\documentclass[11pt]{article}
\usepackage{amsmath}
\usepackage{mathtools}
\usepackage{graphicx}
\usepackage{natbib}
\usepackage{caption}
\usepackage{url} 
\usepackage{enumerate}
\usepackage{bbm}
\usepackage{bm}
\usepackage{amssymb}
\usepackage{verbatim}
\usepackage{subcaption}
\usepackage{extarrows}
\usepackage{algorithm}
\usepackage{algorithmic}
\usepackage{float}
\usepackage[colorlinks=true,citecolor=blue]{hyperref}
\usepackage{geometry}
\usepackage{lscape} 
\usepackage{lipsum}
\usepackage{authblk}
\newcommand{\by}{\boldsymbol{y}}
\newcommand{\blambda}{\boldsymbol{\lambda}}
\newcommand{\btheta}{\boldsymbol{\theta}}
\newcommand{\bbeta}{\boldsymbol{\beta}}
\newcommand{\btau}{\boldsymbol{\tau}}
\newcommand{\bx}{\boldsymbol{x}}

\newcommand{\bgamma}{\boldsymbol{\gamma}}
\DeclareMathOperator*{\argmax}{arg\,max}
\DeclareMathOperator*{\argmin}{arg\,min}

\begin{document}

\title{Monte Carlo Approximation of Bayes Factors via Mixing with Surrogate Distributions}
\author{Chenguang Dai}
\author{Jun S. Liu}
\affil{Department of Statistics, Harvard University}
\maketitle

\begin{abstract}
By mixing the target posterior distribution with a surrogate distribution, of which the normalizing constant is tractable, we propose a  method for estimating the marginal likelihood using the Wang-Landau algorithm. We
show that  a faster convergence of the proposed method can be achieved via the momentum acceleration. Two implementation strategies are detailed: (i) facilitating global jumps between the posterior and surrogate distributions via the Multiple-try Metropolis; (ii) constructing the surrogate via the variational approximation.
When a surrogate is difficult to come by, we describe a new jumping mechanism for general reversible jump Markov chain Monte Carlo algorithms, which combines the Multiple-try Metropolis and a directional sampling algorithm.  We illustrate the proposed methods on several statistical models, including the Log-Gaussian Cox process, the Bayesian Lasso, the logistic regression, and the g-prior Bayesian variable selection.
\end{abstract}

\section{Introduction}
\label{sec:introduction}
Given data $\by$, we consider a finite sequence of competing models $\{\mathcal{M}_k\}$ associated with parameters $\{\btheta_k\}$. The marginal likelihood of data under model $\mathcal{M}_k$, also referred to as the normalizing constant, is defined as
\begin{equation}\nonumber
p(\by\mid \mathcal{M}_k) = \int \gamma(\btheta_k\mid \by, \mathcal{M}_k)  d\btheta_k = \int p(\btheta_k\mid \mathcal{M}_k)p(\by\mid \btheta_k, \mathcal{M}_k)d\btheta_k,
\end{equation}
where $p(\btheta_k\mid\mathcal{M}_k)$ is the prior, $p(\by\mid \btheta_k, \mathcal{M}_k)$ is the likelihood, and $\gamma(\btheta_k\mid\by, \mathcal{M}_k)$ is the unnormalized posterior distribution. To compare different models, Bayesian methods typically compute the Bayes factor, which is defined as the ratio of the normalizing constants under different models, that is, $B_{i, j} = p(\by\mid\mathcal{M}_i)/p(\by\mid\mathcal{M}_j)$. With equal prior probability for models $\mathcal{M}_i$ and $\mathcal{M}_j$, $B_{i, j} > 1$ indicates that model $\mathcal{M}_i$ is more favorable than model $\mathcal{M}_j$ given the current data $\by$. 

We can approximate the Bayes factor by estimating the normalizing constant of each model. For simplicity, we will drop the dependency on $\by$ and the model index $k$ in $\gamma(\btheta_k\mid\by, \mathcal{M}_k)$ when the context is clear, and use $Z_\gamma$ to denote the normalizing constant of $\gamma(\btheta)$. Let $\gamma^\star(\btheta) = \gamma(\btheta)/Z_\gamma$ be the corresponding normalized distribution. Computing $Z_\gamma$ is essentially a task of calculating an integral. However, in many interesting cases, the complex form of the unnormalized density $\gamma(\btheta)$, sometimes with high dimensionality, prohibits us from obtaining either analytic solutions  or easy numerical approximations. Various Monte Carlo strategies have been developed to tackle this problem, such as Chib's method \citep{chib1995marginal}, inverse logistic regression \citep{geyer1994estimating}, importance sampling \citep{gelfand1990sampling}, bridge sampling  \citep{meng1996fitting, meng1996simulating}, path sampling \citep{ogata1989monte}, sequential importance sampling \citep{hammersley1954poor, rosenbluth1955monte, kong&etal94}, and sequential Monte Carlo (SMC) \citep{liu1998sequential, doucet2000sequential, del2006sequential}. A nice overview of nineteen methods for normalizing constant estimation in the context of Bayesian phylogenetics is given in \citet{fourment202019}.

In this article, we present a mixture approach for normalizing constant estimation using the Wang-Landau (WL) algorithm \citep{wang2001efficient}. 
Our idea is to construct a matching surrogate distribution $q(\btheta)$ with its normalizing constant $Z_q$ known, and combine $\gamma(\btheta)$ and $q(\btheta)$   to form a mixture distribution with an adjustable mixing parameter tuned through the WL algorithm. 
The ratio $r = Z_\gamma/Z_q$ is then an easy function of the mixing parameter. Let $q^\star(\btheta)$ be the normalized surrogate distribution. Many of the aforementioned methods also use a surrogate, and the idea of using the WL algorithm to estimate the ratio $r = Z_\gamma/Z_q$ also appears in \cite{liang2005generalized} and \cite{atchade2010wang} in  more restricted settings. 

The proposed WL mixture method is different from existing methods in the following perspectives. First, 
when we apply the WL algorithm in our setting, there is a natural partition of the parameter space indicated by the two (or more if needed) mixture components. Second, unlike the method in \cite{liang2005generalized}, we do not require $\gamma^\star(\btheta)$ and $q^\star(\btheta)$  being well-separated. In fact, jumps between the posterior and the surrogate becomes more flexible if $\gamma^\star(\btheta)$ and $q^\star(\btheta)$ are mixed together.
Third, the WL mixture method does not require $\gamma^\star(\btheta)$ and $q^\star(\btheta)$ to have any overlap. With the help of mode jumping algorithms such as the Multiple-try Metropolis (MTM) \citep{liu2000multiple}, the method is  more robust than importance sampling based methods such as bridge sampling, which crucially rely on the amount of overlaps between $\gamma^\star(\btheta)$ and $q^\star(\btheta)$.

Following \citet{dai2020wang}, we can achieve faster convergence of the proposed method using momentum acceleration. The idea is to formulate the WL algorithm as a (stochastic) gradient descent algorithm minimizing a convex and smooth function, of which the gradient is estimated using Markov chain Monte Carlo (MCMC) samples. Under this optimization framework, we are able to exploit acceleration tools to speed up the convergence of the WL algorithm. Empirically, we find that the simple momentum method improves the efficiency of our algorithm. We illustrate the accelerated WL mixture method on two statistical models, the Log-Gaussian Cox process and the Bayesian Lasso.

Many aforementioned Monte Carlo techniques, including MTM and the WL weight adjustment, are also potentially useful in other Bayesian model selection approaches. In particular, we describe an MTM-based reversible jump MCMC framework, which can be useful 
when it is challenging to propose an appropriate surrogate.
Specifically, we can include the model index $\mathcal{M}_k$ as a parameter in the full posterior distribution specified as
\begin{equation}
\label{eq:multiple-model-comparison-posterior}
p(\btheta_k, \mathcal{M}_k\mid\by) \propto p\left(\by\mid\btheta_k, \mathcal{M}_k\right)p\left(\btheta_k \mid \mathcal{M}_k\right)p\left(\mathcal{M}_k\right),
\end{equation}
and use MCMC to traverse the joint model and parameter space. The ratio between the proportions of time that the Markov chain spends in model $\mathcal{M}_i$ and model $\mathcal{M}_j$, adjusted by the prior on models, consistently estimates the Bayes factor $B_{i,j}$.
A reversible jump MCMC (RJMCMC)  \citep{green1995reversible} algorithm is often required to sample across different dimensional spaces. However, it is well-known that constructing an efficient trans-dimensional proposal is challenging  \citep{brooks2003efficient}.

To enable efficient RJMCMC, we propose to combine MTM and the directional sampling algorithm, 
which will be most effective if $p\left(\btheta_k\mid \by, \mathcal{M}_k\right)$ is uni-modal for each model $\mathcal{M}_k$, and the mode $\widehat{\btheta}_k$ can be located reasonably well beforehand. 
We note that the proposed method is  different from the MTM version of RJMCMC algorithms proposed in \citet{pandolfi2014generalized}. Their method mainly focuses on using a computationally favourable weight function in MTM to avoid evaluating the target density, which can be expensive in complex statistical models. Our method is perhaps most similar to the mode jumping algorithm proposed in \citet{tjelmeland2001mode}. While they design  a mixture of Metropolis-Hastings proposals guided by deterministic local optimization to enable large step-size jumps, we utilize the more flexible MTM.  

The rest of the article is organized as follows. Section \ref{subsec:Wang-Landau-review}   reviews the WL algorithm.
Section \ref{subsec:proposed-method} proposes our mixture formulation, and  explains how we adapt the WL algorithm in the mixture setting to estimate the normalizing constant.
Section \ref{subsec:acceleration} introduces an accelerated version of the WL mixture method.
Section \ref{subsec:VA} describes a principled way of using the variational approximation to construct the surrogate distribution. 
Section \ref{subsec:multiple-try} explains how to use MTM to jump between the two mixture components if $q^\star(\btheta)$ and $\gamma^\star(\btheta)$ are not well aligned and relatively separated. 
Section \ref{subsec:multiple-try-RJMCMC} describes an efficient MTM-RJMCMC algorithm to sample the model space.
Section \ref{sec:comparison} reviews existing methods in the literature, and makes connections and comparisons to our proposed methods. 
Section \ref{sec:illustration} illustrates the utility of the proposed methods on several numerical examples including a Bayesian evaluation of the Log-Gaussian Cox process fitting, a hyper-parameter selection problem for the Bayesian Lasso regression, marginal likelihood estimation for a logistic regression model, 
and  Bayesian variable selection for linear models under the spike-and-slab g-prior.
Section \ref{sec:conclusion} concludes with some final remarks.

\section{A Surrogate Mixture Approach}
\label{sec:methodology} 

\subsection{The Wang-Landau algorithm}
\label{subsec:Wang-Landau-review}
In order to improve the convenience and efficiency of the multicanonical sampling \citep{berg1992multicanonical}, 
\citet{wang2001efficient} proposed a simple stochastic adaptive updating algorithm, which quickly becomes a popular Monte Carlo method for sampling complex physical systems. 
Given a target distribution $p(\btheta)$ and a user-specified partition of the target space $\Theta = \cup_{i = 1}^s \Theta_i$, where $s$ is the total number of subregions, 
we can use the WL algorithm to estimate the probability mass of $p(\btheta)$ within each subregion, i.e., 
$\psi(i) = \mathbbm{P}(\btheta \in \Theta_i)$ for $i\in[s]$, where $[n]$ denotes the set $\{1,\ldots,n\}$ for $\forall n\in\mathbbm{N}_+$.
The main steps of the WL algorithm are outlined in Algorithm \ref{alg:WL-algorithm}.

\begin{algorithm}
\caption{The Wang-Landau algorithm \citep{wang2001efficient}.}
\label{alg:WL-algorithm}
\begin{enumerate}
\item Sample $\btheta_t$ from $K_{t - 1}\left(\btheta_{t - 1}, \cdot\right)$.
\item Update $\psi_t(i) \leftarrow \psi_{t - 1}(i)\left[1 + \eta_t\mathbbm{1}\left(\btheta_t \in \Theta_i\right)\right]$ for $i \in [s]$.
\item Normalize $\{\psi_t(i)\}_{i = 1}^s$ to sum 1. 
\end{enumerate}
\end{algorithm}

To initialize the algorithm, we can simply set $\psi_0(i) = 1/s$, and sample $\btheta_0$ from some initial distribution.
$K_t$ is a Markov kernel invariant to the adaptive target distribution $p^\dagger_t (\btheta)$ defined as
\begin{equation}
\label{eq:twisted-working-target-WL}
p^\dagger_t (\btheta) \propto \sum_{i = 1}^s\frac{p(\btheta)}{\psi_t(i)}\mathbbm{1}(\btheta\in \Theta_i).
\end{equation}
The parameter $\eta_t$ is the learning rate, and typically we shall scale it down following the flat/minimum histogram criterion \citep{zhou2005understanding,belardinelli2007fast} so as to guarantee the convergence of the algorithm. 
The convergence of $\psi_t(i)$ to $\psi(i)$ for $i\in [s]$ has been established in \citet{atchade2010wang} and \citet{fort2015convergence} under proper conditions. Thus, $\btheta_t$ will spend equal amount of time in each subregion $\Theta_i$ as $t \to \infty$. We note that the magnitude of the learning rate $\eta_t$ is informative of the estimation error \citep{zhou2005understanding}. Therefore, a commonly used stopping criteria for the WL algorithm is that $\eta_t$ is small enough.

\subsection{The Wang-Landau mixture method}
\label{subsec:proposed-method}
Suppose we have a (unnormalized) surrogate distribution $q(\btheta)$ 
with its normalizing constant $Z_q$ known. 
In addition, we assume that we have two effective Markov kernels $K_\gamma$ and $K_q$ in hand so that we can sample from $\gamma^\star(\btheta)$ and $q^\star(\btheta)$ sufficiently well. 
Since we control the construction of the surrogate $q^\star(\btheta)$, we can typically make it easy to sample from without using MCMC. More details on constructing the surrogate $q^\star(\btheta)$  are deferred to Section \ref{subsec:VA}. 

The proposed method relies on the following mixture formulation:
\begin{equation}
\pi(\btheta) = \gamma(\btheta) + q(\btheta).
\end{equation}
The key is to 
recognize that the normalizing constants $Z_\gamma$ and $Z_q$ are proportional to the relative probability masses of the two components, $\gamma^\star(\btheta)$ and $q^\star(\btheta)$, in the mixture distribution $\pi(\btheta)$. 
Therefore, we can directly apply the WL algorithm to estimate the ratio $Z_\gamma/Z_q$.\footnote{For numerical stability, we recommend to work on the logarithmic scale.} 
Unlike the standard WL algorithm, which requires a well-separated partition of the parameter space $\Theta$, in our setting the two mixture components $\gamma^\star(\btheta)$ and $q^\star(\btheta)$ naturally defines an overlap-allowed ``partition" of $\Theta$. As detailed below, given $\btheta_t$, we will substitute the deterministic indicator $\mathbbm{1}(\btheta_t \in \Theta_i)$ (see Algorithm \ref{alg:WL-algorithm}) used in the standard WL algorithm by a random indicator $I_t$.

In view of step 1 in Algorithm \ref{alg:WL-algorithm}, it requires an efficient kernel $K_t$ to sample from the adaptive mixture distribution $\pi_t^\dagger(\btheta)$ defined as
\begin{equation}
\pi_t^\dagger(\btheta) \propto \frac{\gamma(\btheta)}{\psi_t(\gamma)} + \frac{q(\btheta)}{\psi_t(q)},
\label{eq:twisted-working-target-constant-WL}
\end{equation}
in which the $\gamma$ and $q$ in the brackets serve as indexes (the same as  $i$ in Algorithm \ref{alg:WL-algorithm}), and should not be misinterpreted as function arguments. In the case where $q^\star(\btheta)$ and $\gamma^\star(\btheta)$ are well mixed, we exploit the data augmentation strategy and perform a Gibbs sampling step \citep{diebolt1994estimation} using $K_\gamma$ and $K_q$. To be specific, we define a binary indicator $I_t$ to denote the mixture component from which $\btheta_t$ comes. Given $\{\psi_{t - 1}(\gamma), \psi_{t - 1}(q)\}$ and $\{\btheta_{t - 1}, I_{t - 1}\}$, if $I_{t - 1} = 1$, we sample $\btheta_t$ from $K_\gamma(\btheta_{t - 1}, \cdot)$, otherwise we sample $\btheta_t$ from $K_q(\btheta_{t - 1}, \cdot)$. Given $\btheta_t$, we then sample $I_t$ from a Bernoulli distribution with probabilities
\begin{equation}
\label{eq:mixture-probability}
\mathbbm{P}(I_t = 1\mid \btheta_t) \propto \gamma(\btheta_t)/\psi_{t - 1}(\gamma),\ \ \ \mathbbm{P}(I_t = 0\mid\btheta_t) \propto q(\btheta_t)/\psi_{t - 1}(q).
\end{equation}
In the case where $q^\star(\btheta)$ and $\gamma^\star(\btheta)$ are not well aligned, we complement Gibbs sampling with the Multiple-try Metropolis (MTM) if informative jumping directions are identified beforehand. We defer a detailed discussion on MTM to Section \ref{subsec:multiple-try}. 

A detailed algorithm implementing the WL mixture method is summarized in Algorithm \ref{alg:WL-constant-algorithm}. The generic Markov kernel $K_t$ invariant to $\pi_t^\dagger(\btheta)$ can be substituted by the Gibbs sampling kernel discussed above or the MTM kernel. In addition, instead of terminating the algorithm after a pre-specified number of iterations $S$, we can also use the stopping criteria that the learning rate $\eta_t$ is small enough (say, below $10^{-3}$).

\begin{algorithm*}
\caption{The Wang-Landau mixture method.}
\label{alg:WL-constant-algorithm}
\begin{enumerate}
\item Algorithmic setup. Choose a decreasing positive sequence $\{\eta_t\}$ as the sequence of learning rate. Set $a_0 = 1$, $c \in (0, 1)$, $\xi_0(\gamma) = \xi_0(q) = 0$, and $\psi_0(\gamma) = \psi_0(q) = 1/2$. Set the total number of iterations to be $S$. We exclude the first $b$ iterations in estimation.
\item At $t = 0$: initialize $\btheta_0$ from some initial distribution, and sample a binary indicator $I_0$ with probability
$\mathbbm{P}(I_0 = 1) \propto \gamma(\btheta_0)$ and $\mathbbm{P}(I_0 = 0) \propto q(\btheta_0)$.
\item For $t\in [S]$: given $(\btheta_{t - 1}, I_{t - 1})$, iterate between the following steps.
\begin{enumerate}
\item Sample $\btheta_t$ from $K_{t - 1}(\btheta_{t - 1}, \cdot)$, which is invariant to the the adaptive mixture distribution $\pi_{t - 1}^\dagger(\btheta)$ defined by $\{\psi_{t - 1}(\gamma), \psi_{t - 1}(q)\}$ as in Equation \eqref{eq:twisted-working-target-constant-WL}.
\item Sample a binary indicator $I_t$ with probabilities
specified in Equation \eqref{eq:mixture-probability}.
\item Update $\{\xi_t(\gamma), \xi_t(q)\}$ and $\{\psi_t(\gamma), \psi_t(q)\}$ as follows:
\begin{equation}
\begin{aligned}
&\xi_t(\gamma) \leftarrow \xi_{t - 1}(\gamma) + \mathbbm{1}\left(I_t = 1\right),\ \ \ \psi_t(\gamma) \leftarrow \psi_{t - 1}(\gamma)\left[1 + \eta_{a_t}\mathbbm{1}\left(I_t = 1\right)\right],\\
&\xi_t(\hspace{0.02cm}q\hspace{0.02cm}) \leftarrow \xi_{t - 1}(\hspace{0.02cm}q\hspace{0.02cm}) + \mathbbm{1}\left(I_t = 0\right),\ \ \ \psi_t(\hspace{0.02cm}q\hspace{0.02cm}) \leftarrow \psi_{t - 1}(\hspace{0.02cm}q\hspace{0.02cm})\left[1 + \eta_{a_t}\mathbbm{1}\left(I_t = 0\right)\right].
\end{aligned}
\end{equation}
\item Normalize $\{\psi_t(\gamma), \psi_t(q)\}$ to sum 1. 
\item If the following condition is satisfied:
\begin{equation}
\label{eq:flatness-criterion}
\frac{\max\{\xi_t(\gamma), \xi_t(q)\}}{\xi_t(\gamma) + \xi_t(q)} - \frac{1}{2} \leq \frac{c}{2},
\end{equation}
update $a_{t + 1} = a_t + 1$ and reset $\xi_t(\gamma) = \xi_t(q) = 0$. Otherwise set $a_{t + 1} = a_t$.
\end{enumerate}
\item Output the estimators $\log\widehat{Z}_\gamma = \log \widehat{r} + \log Z_q$, where $\log\widehat{r} = \frac{1}{S-b}\sum_{t = b + 1}^S[\log\psi_{t}(\gamma) - \log\psi_{t}(q)]$.
\end{enumerate}
\end{algorithm*}

Tunable parameter sequences $\{\xi_t(\gamma),\ \xi_t(q)\}$ are introduced to help check the flat histogram criterion, that is, whether the Markov chain has spent equal amount of time in each of the two mixture components $\gamma^\star(\btheta)$ and $q^\star(\btheta)$. If this is approximately satisfied to the extent controlled by a threshold $c \in (0, 1)$ (see Equation \eqref{eq:flatness-criterion}), we decrease the learning rate and refresh $\xi_t(\gamma) = \xi_t(q) = 0$ so that it can start monitoring the next stage of the algorithm. Empirically we find that the performance of the WL mixture method is robust to the choice of $c$ (see Figure 1 in the supplementary materials). For the numerical examples in the paper, we set $c = 0.2$.

The WL mixture method naturally adapts to the missing data framework. The marginal likelihood of the observed-data can be formulated as 
\begin{equation}\nonumber
L(\by_{\text{obs}}) = \int p(\by_{\text{obs}}\mid\btheta)p(\btheta)d\btheta = \int \int p(\by_{\text{obs}}, \by_{\text{mis}}\mid\btheta)p(\btheta)d\by_{\text{mis}}d\btheta,
\end{equation}
in which $p(\by_{\text{obs}}, \by_{\text{mis}}\mid\btheta)$ is the complete-data distribution, and $p(\btheta)$ is the prior. When the integral $\int p(\by_{\text{obs}}, \by_{\text{mis}}\mid\btheta)d\by_{\text{mis}}$ can be analytically calculated, such as the finite mixture model in which $\by_{\text{mis}}$ are discrete, we can directly apply the WL mixture method to estimate the normalizing constant $L(\by_{\text{obs}})$. More generally, we can treat the 
missing data $\by_{\text{mis}}$ as parameters, and apply the WL mixture method to estimate the normalizing constant of the (unnormalized) complete-data posterior distribution $\gamma(\btheta, \by_{\text{mis}}\mid\by_{\text{obs}}) = p(\by_{\text{obs}}, \by_{\text{mis}}\mid\btheta)p(\btheta)$.

Another extension of the WL mixture method is to introduce a sequence of auxiliary distributions $\{\eta^\star_0(\btheta), \cdots, \eta^\star_p(\btheta)\}$ between the posterior and the surrogate, i.e., $\eta^\star_0(\btheta) = q^\star(\btheta)$ and $\eta^\star_p(\btheta) = \gamma^\star(\btheta)$, and estimate the ratios of the normalizing constants of any two adjacent auxiliary distributions in parallel. This strategy is beneficial if the sequence of auxiliary distributions can be properly chosen.
The idea of constructing auxiliary ``bridging" distributions has been widely used in Monte Carlo simulations, including sequential importance sampling, bridge sampling, and SMC algorithms.
But these earlier methods  typically cannot be easily parallelized. 
Discussions and references on constructing auxiliary distributions are given in Section \ref{subsec:comparison-IS}. We refer to this multiple-step WL mixture method as the parallel WL (PWL) method henceforth, and illustrate it in Section \ref{subsec:log-gaussian-cox} on the Log-Gaussian Cox process, in which we set the prior as the surrogate, and employ a geometric sequence of auxiliary distributions. We believe that the PWL method can complement the WL mixture method when a good surrogate $q^\star(\btheta)$ is not easy to construct.

\subsection{Acceleration of the Wang-Landau mixture method}
\label{subsec:acceleration}
The efficiency of the WL mixture method can be further improved using the acceleration idea discussed in \citet{dai2020wang}. The update of the reweighting factor $\psi_t(\gamma)$ (step 3(c) in Algorithm \ref{alg:WL-constant-algorithm}) can be approximated as follows:
\begin{equation}
\label{eq:WL-update-approximation}
\begin{aligned}
\log\psi_t(\gamma) & = \log\psi_{t - 1}(\gamma) + \log\left[1 + \eta_{a_t}\mathbbm{1}\left(I_t = 1\right)\right]\\
& \approx \log\psi_{t - 1}(\gamma) + \eta_{a_t}\mathbbm{1}\left(I_t = 1\right).
\end{aligned}
\end{equation}
We note that the above approximation is fairly accurate since the learning rate $\eta_t
$ is small.

Let $u_t(\gamma) = \log\psi_t(\gamma)$ and $u_t(q) = \log\psi_t(q)$. We consider the following optimization problem:
\begin{equation}
\label{eq:optimization}
\begin{aligned}
& \min_{u_1, u_2\in\mathbbm{R}}f(u_1, u_2) = \log\left(\exp(u^\star(\gamma) - u_1) + \exp(u^\star(q) - u_2)\right),\\
&\text{subject to}\ \ u_1 + u_2 = 0,
\end{aligned}
\end{equation}
in which $u^\star(\gamma) = \log Z_\gamma - \log Z_q$ and $u^\star(q) = \log Z_q - \log Z_\gamma$. It is not difficult to see that this is a convex optimization problem because the objective function $f(u_1, u_2)$ is a log-sum-exp function and the constraint is linear. It has a unique solution at $(u^\star(\gamma), u^\star(q))$.

To solve the constrained optimization problem \eqref{eq:optimization}, we use the projected gradient descent algorithm. More precisely, one-step update of $u_t(\gamma)$ is
\begin{equation}
\begin{aligned}
u_t(\gamma) & = u_{t - 1}(\gamma) - \eta_{a_t}\frac{\partial f}{\partial u_1}\left(u_{t - 1}(\gamma), u_{t - 1}(q)\right)\\
& = u_{t - 1}(\gamma) + \eta_{a_t}\mathbbm{P}\left(I_t = 1\right),
\end{aligned}
\end{equation}
in which $\mathbbm{P}(I_t = 1)$ equals the weight of $\gamma^\star(\btheta)$ in the mixture distribution $\pi^\dagger_{t - 1}(\btheta)$ (see Equation \eqref{eq:twisted-working-target-constant-WL}). We update $u_t(q)$ in a similar way. Note that the analytical evaluation of $\mathbbm{P}\left(I_t = 1\right)$ involves the unknown normalizing constant $Z_\gamma$, and is thus infeasible in practice. However, we can implement an
one-step or multiple-step Monte Carlo approximation to $\mathbbm{P}\left(I_t = 1\right)$, e.g., by $\mathbbm{1}\left(I_t = 1\right)$. This recovers the WL update in Equation \eqref{eq:WL-update-approximation}. In addition, the projection step to the set $\{u_1, u_2\in\mathbbm{R}, u_1 + u_2 = 0\}$ is equivalent to the normalization step (see step 3(d) in Algorithm \ref{alg:WL-constant-algorithm}). Therefore, the WL algorithm is equivalent to the stochastic projected gradient descent algorithm solving the constrained optimization problem \eqref{eq:optimization}.

Once we have the optimization perspective, various acceleration tools can be employed to improve the efficiency of the WL mixture method. One simple tool we find useful is the momentum method, which exponentially accumulates a momentum vector to amplify the persistent gradient across iterations, thus reducing the oscillation caused by the noise in the gradient estimate. More precisely, we define a momentum vector $(m_t(\gamma), m_t(q))$ with initialization $m_0(\gamma) = m_0(\hspace{0.02cm}q\hspace{0.02cm}) = 0$, and modify step 3(c) in Algorithm \ref{alg:WL-constant-algorithm} as below.

\vspace{2mm}

\indent 3 (c$^\prime$) (Momentum accelerated WL updates)

\vspace{1.5mm}

    \indent\indent(i) Update the momentum vector:
    \begin{equation}\nonumber
     m_t(\gamma) \leftarrow \beta m_{t - 1}(\gamma) - \eta_{a_t}\mathbbm{1}\left(I_t = 1\right),\ \ \ \ m_t(\hspace{0.02cm}q\hspace{0.02cm}) \leftarrow \beta m_{t - 1}(\hspace{0.02cm}q\hspace{0.02cm}) - \eta_{a_t}\mathbbm{1}\left(I_t = 0\right).
    \end{equation}
    \indent\indent(ii) Update the reweighting vector:
    \begin{equation}\nonumber
    \log\psi_t(\gamma) \leftarrow \log\psi_{t - 1}(\gamma) - m_t(\gamma),\ \ \ \ \ \ \ \log\psi_t(\hspace{0.02cm}q\hspace{0.02cm}) \leftarrow \log\psi_{t - 1}(\hspace{0.02cm}q\hspace{0.02cm}) - m_t(\hspace{0.02cm}q\hspace{0.02cm}).
    \end{equation}
    \indent\indent(iii) Update $\{\xi_t(\gamma), \xi_t(q)\}$ as in step 3(c) in Algorithm \ref{alg:WL-constant-algorithm}.
    
\vspace{2mm}

The tuning parameter $\beta$ is commonly set to be 0.9 or higher, which calibrates the fraction of the accumulated past gradients that we want to incorporate into the current update. Numerical illustrations of the accelerated WL mixture method is given in Figure \ref{fig:AccWL} on two statistical models, the Log-Gaussian Cox process and the Bayesian Lasso.

\subsection{Constructing the surrogate distribution}
\label{subsec:VA}
In principle, any posterior approximation with a known normalizing constant, such as the Laplace approximation and the variational approximation, can be used to construct the surrogate distribution. We can also use MCMC methods to obtain posterior samples, and fit some parametric distribution to them. In this section, we describe how to construct a surrogate $q(\btheta)$ using the variational approximation  \citep{jordan1999introduction,blei2017variational}. The variational approach enjoys two main advantages. First, it is computationally efficient and does not require MCMC sampling to explore $\gamma^\star(\btheta)$. Second, it provides a reasonable approximation to $\gamma^\star(\btheta)$ in a wide class of statistical models \citep{wainwright2008graphical}. 

The variational approximation aims at finding the closest distribution $q^\star(\btheta)$ to $\gamma^\star(\btheta)$ in the KL divergence within a particular class of distributions $Q$, that is,
\begin{equation}
\label{eq:VI-KL}
q^\star(\btheta) = \argmin_{p(\btheta) \in Q} \text{KL}\left(p(\btheta) || \gamma^\star(\btheta)\right).
\end{equation}
$\text{KL}\left(p(\btheta) || \gamma^\star(\btheta)\right)$ is not computable as it involves the unknown normalizing constant $Z_\gamma$. However, we can equivalently reformulate the optimization problem \eqref{eq:VI-KL} as follows:
\begin{equation}
\label{eq:VI-KL-equivalent}
q^\star(\btheta) = \argmax_{p(\btheta) \in Q} \text{ELBO}(p) = \argmax_{p(\btheta) \in Q} \Big\{\mathbbm{E}_{p}\left[\log \gamma(\btheta) \right] - \mathbbm{E}_{p}\left[\log p(\btheta) \right]\Big\},
\end{equation}
in which $Z_\gamma$ is no longer involved. ELBO refers to the \textit{evidence lower bound} of $\log Z_\gamma$ since
\begin{equation}
\label{eq:VI-ELBO}
\log Z_\gamma = \text{KL}\left(q^\star(\btheta) || \gamma^\star(\btheta)\right) + \text{ELBO}(q^\star) \geq \text{ELBO}(q^\star).
\end{equation}
We note that the EM algorithm \citep{dempster1977maximum} can also be formulated as a two-step iterative algorithm
maximizing the ELBO with respect to $p(\btheta)$ and the relevant model parameters \citep{tzikas2008variational}. 

Before solving the optimization problem \eqref{eq:VI-KL-equivalent}, we need to specify the variational family $Q$. A commonly considered class of distributions $Q$ is the mean-field variational family, which assumes that $q^\star(\btheta)$ is a product of univariate distributions, that is, $q^\star(\btheta) = \prod_{j = 1}^d q_j^\star(\theta_j)$. We assume that $q_j^\star(\theta_j)$ belongs to some parametric family $Q_j$ whose probability density function can be evaluated exactly. 

To solve the optimization problem \eqref{eq:VI-KL-equivalent}, we can use the \textit{coordinate ascent variational inference} (CAVI) algorithm \citep{bishop2006pattern}. CAVI, detailed in Algorithm \ref{alg:CAVI}, iteratively maximizes the ELBO in a coordinate-wise fashion.
We note that the optimization problem in \eqref{eq:VI-ELBO-coordinate-wise} can be further simplified in conjugate cases. For each $j \in [d]$, conditioning on all the other components $q_i^\star(\theta_i),\ i \neq j$, $\text{ELBO}(q_j)$ can be rewritten as
\begin{equation}
\label{eq:VI-ELBO-coordinate-wise-conjugate}
\text{ELBO}(q_j) = - \text{KL}\left(q_j(\theta_j) || q^{\text{opt}}_j(\theta_j)\right) + \text{constant}.
\end{equation}
$q^{\text{opt}}_j(\theta_j) \propto \exp\left[\mathbbm{E}_{-j}\left(\log \gamma(\theta_j, \btheta_{-j})\right)\right]$, where $\mathbbm{E}_{-j}$ is taken with respect to the density $\prod_{i\neq j}q_i^\star(\theta_i)$. If $q^{\text{opt}}_j(\theta_j) \in Q_j$ (conjugacy), the optimal $q_j(\theta_j)$ is $q^{\text{opt}}_j(\theta_j)$ since the KL divergence is non-negative. 

\begin{algorithm*}
\caption{The coordinate ascent variational inference (CAVI) algorithm \citep{blei2017variational}.}
\label{alg:CAVI} 
\begin{enumerate}
\item Initialize each $q_j^\star(\theta_j) \in Q_j$ for $j \in [d]$.
\item For each $j \in [d]$, fix all the other components $q_i^\star(\theta_i),\ i \neq j$, update $q_j^\star(\theta_j)$ with 
\begin{equation}
\label{eq:VI-ELBO-coordinate-wise}
q_j^\star(\theta_j) = \argmax_{q_j \in Q_j}\text{ELBO}(q_j) =\argmax_{q_j \in Q_j}\Big\{ \mathbbm{E}_j\big[\mathbbm{E}_{-j}\left(\log \gamma(\theta_j, \btheta_{-j})\right)\big] - \mathbbm{E}_j\left[\log q_j(\theta_j)\right]\Big\},
\end{equation}
where $\mathbbm{E}_j$ and $\mathbbm{E}_{-j}$ are taken with respect to the densities $q_j(\theta_j)$ and $\prod_{i\neq j}q_i^\star(\theta_i)$, respectively.
\item Calculate $\text{ELBO}(q^\star)$ with $q^\star(\btheta) = \prod_{j = 1}^d q_j^\star(\theta_j)$. If ELBO has not converged, go back to step 2. Otherwise output $q^\star(\btheta)$. 
\end{enumerate}
\end{algorithm*}

\section{Global jump via Multiple-try Metropolis}
\label{sec:multiple-try}
\subsection{The Multiple-try Metropolis}
\label{subsec:multiple-try}
In many problems, the two mixture components $q^\star(\btheta)$ and  $\gamma^\star(\btheta)$ 
may not be well aligned and sometimes can be completely separated. Thus,  naive Metropolis-Hastings proposals may be easily  trapped in one of the components and cannot efficiently traverse the whole space. We here describe an approach based on 
the Multiple-try Metropolis (MTM) method \citep{liu2000multiple} for constructing a proper
Markov kernel  that enables easy jumps between different modes.
 
Given a target distribution $\pi(\bx)$ defined on $\mathbbm{R}^d$ and a proposal transition function $T(\bx, \by)$, a version of MTM is described in Algorithm \ref{alg:mtm_standard}. Heuristically, MTM  aims at \textit{biasing}  the multiple proposals with a proper weight function $w(\bx,\by)$:
\begin{equation}
w(\bx, \by) = \pi(\bx)T(\bx, \by)\lambda(\bx,\by),
\end{equation}
in which $\lambda(\bx,\by)$ is a user-chosen nonnegative symmetric function. 
More precisely, given the current state $\bx_t$, MTM draws $m$ proposals $\{\by^{(1)}, \cdots, \by^{(m)}\}$ from the transition function $T(\bx_t, \cdot)$, and then 
selects $\by$ from $\{\by^{(1)}, \cdots, \by^{(m)}\}$ with probability proportional to $w(\by^{(j)}, \bx_t)$. A proper acceptance-rejection rule (steps 3 and 4 in Algorithm \ref{alg:mtm_standard}) 
is employed  to ensure the reversibility of the Markov chain. 
 
\begin{algorithm*}
\caption{The Multiple-try Metropolis \citep{liu2000multiple}.}
\label{alg:mtm_standard}
\begin{enumerate}
\item Sample $\by^{(1)}, \cdots, \by^{(m)}$ i.i.d from $T(\bx_t, \cdot)$. Compute the weight function $w(\by^{(j)}, \bx_t)$.
\item Sample $\by$ from $\by^{(1)}, \cdots, \by^{(m)}$ with probability proportional to $w(\by^{(j)}, \bx_t)$.  
\item Given $\by$, sample $\bx^{(1)}, \cdots, \bx^{(m - 1)}$ i.i.d from $T(\by, \cdot)$. Set $\bx^{(m)}= \bx_t$.
\item Accept $\bx_{t + 1} = \by$ with probability:
\begin{equation}
\alpha = \min\bigg\{1,\ \ \ \frac{w(\by^{(1)}, \bx_t) + \cdots + w(\by^{(m)}, \bx_t)}{w(\bx^{(1)}, \by) + \cdots + w(\bx^{(m)}, \by)}\bigg\}.
\end{equation}
\end{enumerate}
\end{algorithm*}

A special choice of $\lambda$ is $\lambda(\bx, \by) = [T(\bx, \by) + T(\by, \bx)]^{-1}$. If $T(\bx,\by)$ is also a symmetric proposal, the corresponding acceptance probability simplifies to:
\begin{equation}
\alpha = \min\bigg\{1,\ \ \ \frac{\pi(\by^{(1)}) + \cdots + \pi(\by^{(m)})}{\pi(\bx^{(1)}) + \cdots + \pi(\bx^{(m)})}\bigg\}.
\end{equation}
This special case  is referred to as MTM (II)  in \cite{liu2000multiple}. MTM  is particularly useful when it is combined with  a directional sampling algorithm. For instance, 
if we know a desirable jumping direction, we can use MTM to explore a wide range along it. Let $\bm{e}$ denote the jumping direction. For the simple case where $\bm{e}$ is fixed and independent of the current state $\bx_t$, we outline the main steps in Algorithm \ref{alg:multiple-try-direction-sampling}. More generally, we can choose the jumping direction $\bm{e}$ based on $\bx_t$. Some detailed discussion on a special form of this adaptive strategy can be found in Section \ref{subsec:multiple-try-RJMCMC}.

\begin{algorithm}
\caption{The Multiple-try Metropolis combined with the directional sampling algorithm.}
\label{alg:multiple-try-direction-sampling} 
\begin{enumerate}
\item Sample $r^{(1)}, \cdots, r^{(m)}$ from a user-chosen distribution $p(r)$. Let $\by^{(j)} = \bx_t + r^{(j)} \cdot \bm{e}$. Compute the target density $\pi(\by^{(j)})$.
\item Sample $\by$ from $\by^{(1)}, \cdots, \by^{(m)}$ with probability proportional to $\pi(\by^{(j)})$. Set $\bx^{(j)} = \by - r^{(j)} \cdot \bm{e}$.  
\item Accept $\bx_{t + 1} = \by$ with probability:
\begin{equation}
\alpha = \min\bigg\{1,\ \ \ \frac{\pi(\by^{(1)}) + \cdots + \pi(\by^{(m)})}{\pi(\bx^{(1)}) + \cdots + \pi(\bx^{(m)})}\bigg\}.
\end{equation}
\end{enumerate}
\end{algorithm}

When we incorporate MTM in the WL mixture method, assuming that we are equipped with an efficient kernel $K_\gamma$, some pre-MCMC runs should help us pin down informative jumping directions, such as the directions connecting the modes of the posterior and the surrogate. Thus, 
we can substitute step 3(a) in Algorithm \ref{alg:WL-constant-algorithm} by randomly alternating between MTM and Gibbs sampling.

\begin{figure*}
\centering
\begin{subfigure}[t]{0.46\columnwidth}
\centering
\includegraphics[width=0.8\columnwidth]{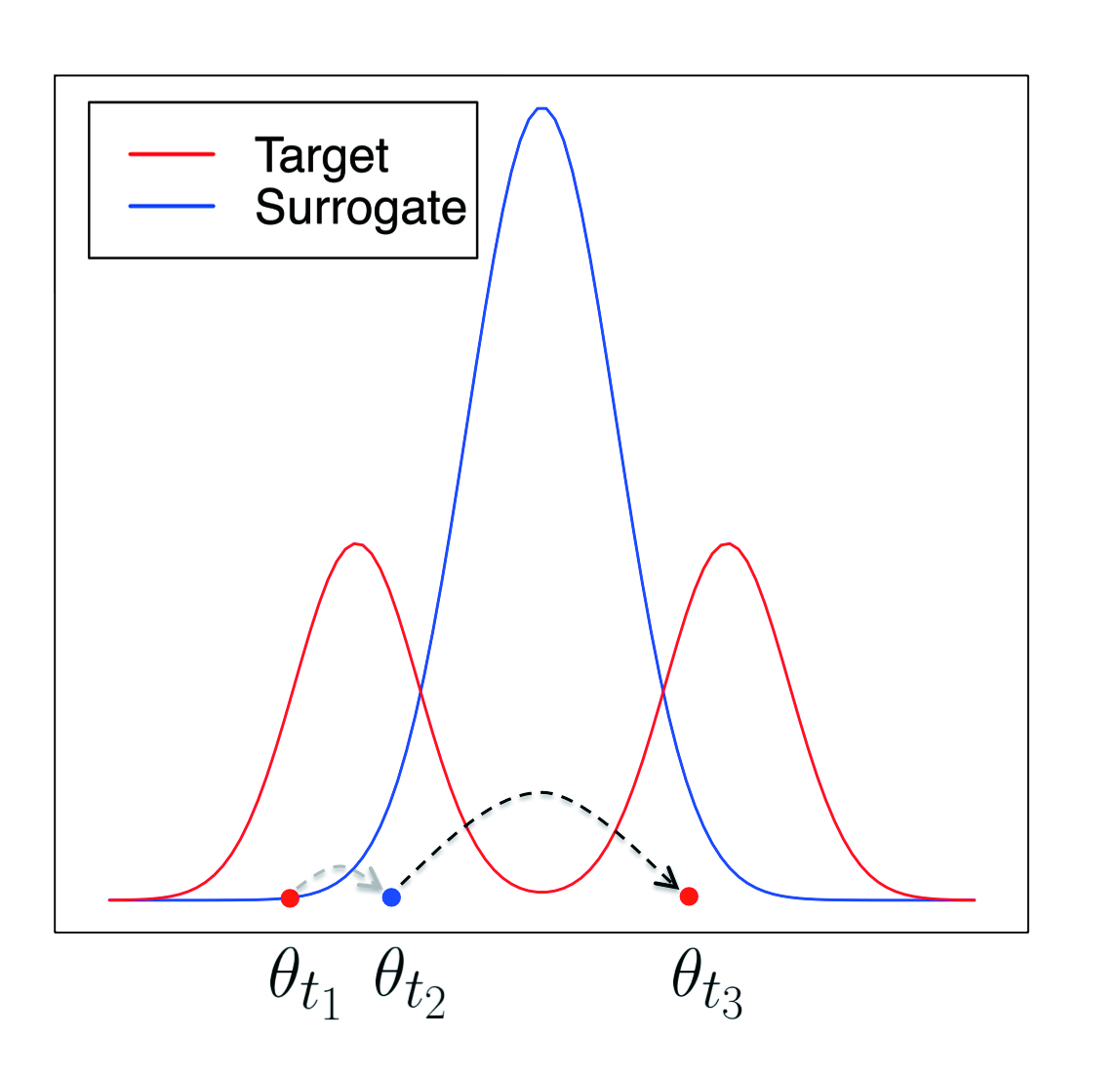}
\caption{Demonstration of mode jumping. The color indicates the mixture component that the sample comes from. 
Initially, $\btheta_{t_1}$ is considered to be more likely a sample from $\gamma^\star(\btheta)$ (step 3(b) in Algorithm \ref{alg:WL-constant-algorithm}). The WL mixture method keeps downweighting the mixture component $\gamma^\star(\btheta)$ by increasing its reweighting factor $\psi_{t_1}(\gamma)$ (step 3(c) in Algorithm \ref{alg:WL-constant-algorithm}). In the meantime, we perform Gibbs sampling steps, which locally moves the chain using the Markov kernel $K_\gamma$, thus the chain still stays around the same local mode. After $\gamma^\star(\btheta)$ has been downweighted enough, at some point $t_2$,  $\btheta_{t_2}$ is considered to be more likely a sample from $q^\star(\btheta)$. In the next step we directly sample from $q^\star(\btheta)$, leading to a global jump from $\btheta_{t_2}$ to $\btheta_{t_3}$, which locates the other mode of $\gamma^\star(\btheta)$.}
\label{subfig:demo-multimodality}
\end{subfigure}
\hspace{0.8cm}
\begin{subfigure}[t]{0.46\columnwidth}
\centering
\includegraphics[width=0.8\columnwidth]{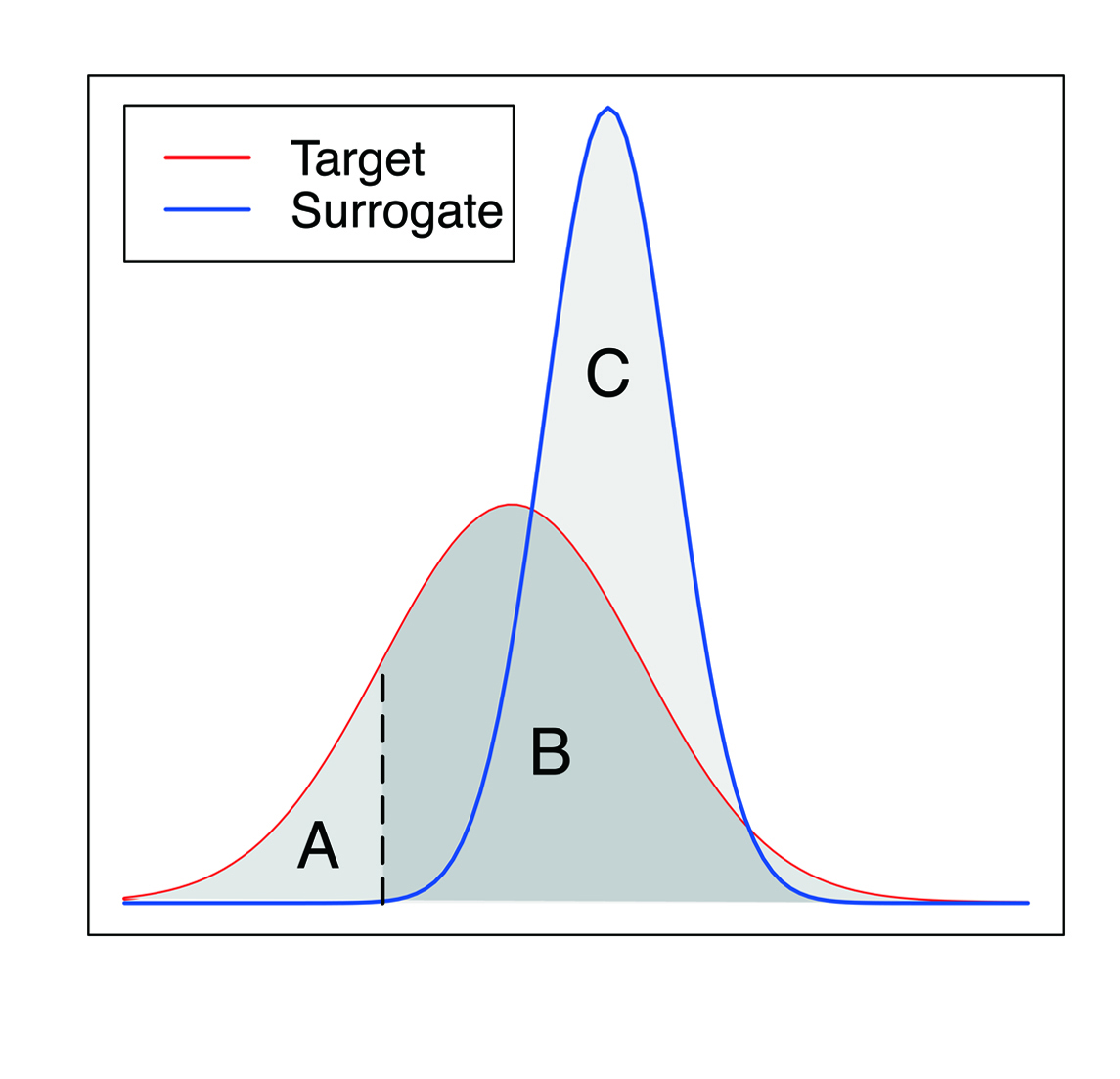}
\caption{Comparison between the WL mixture method and importance sampling. Importance sampling only estimates the normalizing constant of the target distribution restricted on the region B, thus yields an underestimated normalizing constant. 
In contrast, the WL mixture method has approximately equal chance to explore the whole high-density regions of the target and the surrogate distributions, thus produces more accurate normalizing constant estimates.}
\label{subfig:demo-WL}
\end{subfigure}%
\caption{}
\end{figure*}

When the posterior $\gamma^\star(\btheta)$ is multimodal, the proposed surrogate mixture framework can potentially help identify the multimodality of $\gamma^\star(\btheta)$,
upon which a better $K_\gamma$ can be designed using MTM. 
The idea is similar to parallel tempering \citep{geyer1991}, which relies on a sequence of auxiliary distributions so that global jumps are possible by ``transporting'' samples back and forth from the posterior to  auxiliary distributions. In our setting, the surrogate $q^\star(\btheta)$ plays a similar role as the auxiliary distributions used in parallel tempering, and the WL weight adjustment ensures that transitions between the posterior and the auxiliary distribution are sufficiently frequent.
As shown in Figure \ref{subfig:demo-multimodality}, we consider a setting where $\gamma^\star(\btheta)$ is bimodal and $q^\star(\btheta)$ covers both modes, but we are unaware of the multimodality of $\gamma^\star(\btheta)$ in the first place and $K_\gamma$ only enables local moves. As is explained in the caption of Figure \ref{subfig:demo-multimodality}, by leveraging the surrogate $q^\star(\btheta)$ under the help of the WL reweighting, 
the algorithm helps the chain jump across the two modes of $\gamma^\star(\btheta)$.

\subsection{An extension: MTM for reversible jump MCMC}
\label{subsec:multiple-try-RJMCMC}
We discuss possible utilizations of some aforementioned ideas in the classical reversible-jump MCMC (RJMCMC) \citep{green1995reversible} framework for Bayesian model comparison. That is, 
instead of estimating the normalizing constant of each model, we incorporate the model index into the joint posterior distribution defined in Equation \eqref{eq:multiple-model-comparison-posterior}, and sample the model indicator and model parameters simultaneously. 
It generally requires RJMCMC since the posterior distribution $p\left(\btheta_k, \mathcal{M}_k\mid \by\right)$ is potentially trans-dimensional.  

Effective jumping mechanisms are necessary for successful traverse across different model spaces. We believe both MTM and the WL weight adjustment are potentially useful in constructing efficient trans-dimensional proposals. (i) The MTM directional sampling may guide the Markov chain to directly jump towards the mode of the within-model posterior, so that the acceptance probability can be much higher than other generic jumping mechanisms. (ii) The WL weight adjustment can be used to balance the probability masses of different models in the joint posterior distribution $p\left(\btheta_k, \mathcal{M}_k\mid \by\right)$, so as to facilitate trans-dimensional jumps. In this paper, we detail the discussion on the MTM directional sampling in RJMCMC, and leave the idea of using the WL weight adjustment in RJMCMC for future explorations.

We note that the MTM-based reversible jump algorithm (MTM-RJMCMC) is most useful if (i) $p(\btheta_k \mid  \by, \mathcal{M}_k)$ is approximately unimodal for each model  $\mathcal{M}_k$; (ii) we can estimate the mode $\widehat{\btheta}_k$ reasonably well before running the algorithm. 
For $i \neq j$, suppose we want to move from model $\mathcal{M}_i$ to model $\mathcal{M}_j$. Since $\btheta_i\in\mathbbm{R}^{d_i}$ and $\btheta_j\in\mathbbm{R}^{d_j}$ are potentially in different dimensions, we first match the dimensions of $\btheta_i$ and $\btheta_j$ by introducing auxiliary parameters $\bm{u} \in \mathbbm{R}^{d_j}$ and $\bm{v} \in \mathbbm{R}^{d_i}$ so that the dimension and domain of $(\btheta_i, \bm{u})$ match those of $(\bm{v}, \btheta_j)$. We note that this is just one principled way to match the parameter spaces. For specific problems, more efficient designs may exist and should be considered. 

We define the augmented posterior distributions as
\begin{equation}
\label{eq:augment-posterior-multiple-model}
\begin{aligned}
&p_i\left(\btheta_i, \bm{u}, \mathcal{M}_i \mid \by\right) = p(\btheta_i\mid \by, \mathcal{M}_i)q_i(\bm{u})p(\mathcal{M}_i), \\
&p_j\left(\bm{v}, \btheta_j, \mathcal{M}_j\mid \by \right) =p(\btheta_j\mid \by, \mathcal{M}_j)q_j(\bm{v})p(\mathcal{M}_j),
\end{aligned}
\end{equation}
in which $q_i(\bm{u})$ and $q_j(\bm{v})$ are user-chosen unimodal distributions with modes denoted as $\widehat{\bm{u}}$ and $\widehat{\bm{v}}$. The above construction implies that  $\btheta_i \perp \bm{u}$ and $\btheta_j \perp \bm{v}$. In general, we can consider introducing dependence structures between $\btheta_i, \bm{u}$ and $\btheta_j, \bm{v}$. 

The multiple-try trans-dimensional move from model $\mathcal{M}_i$ to model $\mathcal{M}_j$ is summarized in Algorithm \ref{alg:mtm-RJMCMC} and briefly explained here. Given the current state $\btheta_i$, we first sample $\bm{u}$ from $q_i(\bm{u})$, and then construct $m$ proposals $(\bm{v}^{(k)}, \btheta_j^{(k)}) = (\btheta_i, \bm{u}) + r^{(k)}\cdot \bm{e}$ for $k \in [m]$. Two types of directional jumping mechanisms can be exploited: (i) fixed-directional jump; (ii) adaptive-directional jump. For the fixed-directional jump, the jumping direction is defined by the two pre-located modes of the augmented posteriors $p_i\left(\btheta_i, \bm{u}, \mathcal{M}_i \mid \by\right)$ and $p_j\left(\bm{v}, \btheta_j, \mathcal{M}_j\mid \by \right)$, and is fixed throughout the algorithm. For the adaptive-directional jump, the jumping direction is defined by the current state of the chain $(\btheta_i, \bm{u})$ and the mode of the augmented posterior $p_j\left(\bm{v}, \btheta_j, \mathcal{M}_j\mid \by \right)$.

There are subtle differences in the implementation of the two jumping mechanisms. If we use the adaptive-directional jump to {\it jump towards} a mode, that is,  $\bm{e} = (\widehat{\bm{v}} - \btheta_i, \widehat{\btheta}_j - \bm{u})/||(\widehat{\bm{v}} - \btheta_i, \widehat{\btheta}_j - \bm{u})||$, the sampling distribution $p(r)$ of the jumping distance $r$ is required to be a centered symmetric distribution in order that the acceptance probability can be simplified as in Equation (\ref{eq:accept-probability-mtm-RJMCMC-adaptive-direction}). A more general $p(r)$ is allowed if we use a generalized form of MTM in \cite{liu2000multiple}.
In contrast, for the fixed-directional jump, we can simply set the jumping direction as $\bm{e} = (\widehat{\bm{v}} - \widehat{\btheta}_i, \widehat{\btheta}_j - \widehat{\bm{u}})$ without standardization, and sample the jumping distance $r$ from an arbitrary distribution, not necessarily being symmetric and centered at 0. In fact, to push the chain directly jump into the mode $(\widehat{\bm{v}}, \widehat{\btheta}_j)$, we recommend to center $p(r)$ at 1. We note that the acceptance probability of the adaptive-directional jump involves an additional Jacobian defined as
\begin{equation}
\label{eq:adaptive-directional-jump-Jacobian}
J((\btheta_i, \bm{u}), (\bm{v}, \btheta_j)) = \left|1 - \frac{||(\bm{v} - \btheta_i, \btheta_j - \bm{u})||}{||(\widehat{\bm{v}} - \btheta_i, \widehat{\btheta}_j - \bm{u})||}\right|^{d_i + d_j - 1}.
\end{equation}
For the fixed-directional jump, the Jacobian is not required.

Each of the two jumping mechanisms has its own advantages depending on the scenarios. For instance, if the local variations around two modes ($\widehat{\btheta_i}, \widehat{\bm{u}})$ and $(\widehat{\bm{v}}, \widehat{\btheta_j})$ differ significantly, the adaptive-directional jump is more favorable, as the fixed jumping direction can be misleading when the chain jumps from the relatively wider mode to the narrower one. On the other hand, since 
$p(r)$ for the fixed-directional jump is more flexible, e.g., $p(r)$ can be centered at 1, when the jumping direction $\bm{e} = (\widehat{\bm{v}} - \widehat{\btheta}_i, \widehat{\btheta}_j - \widehat{\bm{u}})$ is indeed informative, it appears more efficient than the adaptive-directional jump where $p(r)$ centers at 0.

We then sample $(\bm{v}, \btheta_j)$ from the multiple proposals $\{(\bm{v}^{(k)}, \btheta_j^{(k)})\}_{k = 1}^m$, with probability proportional to the augmented posterior density $p_j(\bm{v}^{(k)}, \btheta_j^{(k)}, \mathcal{M}_j \mid \by)$. After obtaining $(\bm{v}, \btheta_j)$, we set $(\btheta_i^{(k)}, \bm{u}^{(k)}) = (\bm{v}, \btheta_j) - r^{(k)}\cdot \bm{e}$ for $k\in[m]$. We accept the trans-dimensional proposal $(\bm{v}, \btheta_j)$ with probability $\alpha$ given in Equations \eqref{eq:accept-probability-mtm-RJMCMC-fixed-direction} and \eqref{eq:accept-probability-mtm-RJMCMC-adaptive-direction}, depending on which jumping mechanism we use. We note that the proposed trans-dimensional move should be combined with local MCMC moves within each model $\mathcal{M}_k$. Since the current setting is slightly different from that of a typical MTM, we provide here a theoretical justification in the following proposition. The proof of Proposition 1 can be found in the supplementary materials. \\

\begin{algorithm*}
\caption{The Multiple-try Metropolis reversible jump MCMC algorithm.}
\label{alg:mtm-RJMCMC}
\vspace{0.3cm}
For $i \neq j$, suppose the current posterior  draw $\btheta_i$ is from
model $\mathcal{M}_i$, the trans-dimensional move  to model $\mathcal{M}_j$ is accomplished as follows.
\begin{enumerate}
\item Sample the auxiliary variable $\bm{u}$ from $q_i(\bm{u})$, of which the dimension and domain match those of $\btheta_j$ in model $\mathcal{M}_j$.
\item Set the jumping direction and sample the jumping distances.
\begin{enumerate}
\item (Fixed-directional jump) Set the jumping direction as $\bm{e} = (\widehat{\bm{v}} - \widehat{\btheta}_i, \widehat{\btheta}_j - \widehat{\bm{u}})$. Sample the jumping distances $r^{(1)}, \cdots, r^{(m)}$ from an arbitrary distribution $p(r)$ (recommend to center $p(r)$ at 1).
\item (Adaptive-directional jump) Set the jumping direction as $\bm{e} = (\widehat{\bm{v}} - \btheta_i, \widehat{\btheta}_j - \bm{u})/||(\widehat{\bm{v}} - \btheta_i, \widehat{\btheta}_j - \bm{u})||$. Sample the jumping distances $r^{(1)}, \cdots, r^{(m)}$ from a \textit{symmetric} distribution $p(r)$ centered at 0. 
\end{enumerate}
\item Propose multiple tries: set $(\bm{v}^{(k)}, \btheta_j^{(k)}) = (\btheta_i, \bm{u}) + r^{(k)}\cdot \bm{e}$ for $k\in[m]$.
\item Sample $(\bm{v}, \btheta_j)$ from $\{(\bm{v}^{(k)}, \btheta_j^{(k)})\}_{k = 1}^m$ with probability proportional to $p_j(\bm{v}^{(k)}, \btheta_j^{(k)}, \mathcal{M}_j \mid \by)$.
\item Given $(\bm{v}, \btheta_j)$, set $(\btheta_i^{(k)}, \bm{u}^{(k)}) = (\bm{v}, \btheta_j) - r^{(k)}\cdot \bm{e}$ for $k\in[m]$.
\item Accept $(\bm{v}, \btheta_j)$ with probability $\alpha$ specified as below.
\begin{enumerate}
\item (Fixed-directional jump)
\begin{equation}
\label{eq:accept-probability-mtm-RJMCMC-fixed-direction}
\alpha = \min\Bigg\{1,\ \ \ \frac{\sum_{k = 1}^m p_j(\bm{v}^{(k)}, \btheta_j^{(k)}, \mathcal{M}_j \mid \by)}{\sum_{k = 1}^m p_i(\btheta_i^{(k)}, \bm{u}^{(k)}, \mathcal{M}_i \mid \by)}\Bigg\}.
\end{equation}
\item (Adaptive-directional jump)
\begin{equation}
\label{eq:accept-probability-mtm-RJMCMC-adaptive-direction}
\alpha = \min\Bigg\{1,\ \ \ \frac{\sum_{k = 1}^m p_j(\bm{v}^{(k)}, \btheta_j^{(k)}, \mathcal{M}_j \mid \by)}{\sum_{k = 1}^m p_i(\btheta_i^{(k)}, \bm{u}^{(k)}, \mathcal{M}_i \mid \by)}\times J((\btheta_i, \bm{u}), (\bm{v}, \btheta_j))\Bigg\}.
\end{equation}
\end{enumerate}
\end{enumerate}
\end{algorithm*}

\noindent\textbf{Proposition 1:} The proposed trans-dimensional move, equipped with either the fixed-directional jumping mechanism or the adaptive-directional jumping mechanism, leaves the posterior distribution $p\left(\btheta_k, \mathcal{M}_k \mid \by\right)$ invariant.

\section{Review of Existing Methods with Comparisons}
\label{sec:comparison}
\subsection{Importance sampling and sequential Monte Carlo}
\label{subsec:comparison-IS}
It is known that the performance of importance sampling is determined by how closely 
 the proposal distribution tracks the target distribution. In a good importance sampler, high probability regions of the proposal and target distributions overlap substantially, and the proposal typically has a heavier tail than the target. Otherwise,  the variance of the importance sampling estimator can be unacceptably large so that the resulting estimation is misleading. 
If the dimension of the problem is high, it is generally hard to construct an appropriate proposal distribution.

Figure \ref{subfig:demo-WL} provides a cartoon illustration 
of the setting that the surrogate distribution has a smaller domain (and thinner tail) compared to the target distribution. In this case, importance sampling is likely to underestimate the normalizing constant (see the caption in Figure \ref{subfig:demo-WL}).
This phenomenon is illustrated on two realistic examples in Section \ref{sec:illustration}, the Log-Gaussian Cox process and the Bayesian Lasso. 

Sequential Monte Carlo (SMC), developed upon importance sampling, employs a sequential scheme to handle the high-dimensionality of the target distribution \citep{liu2001theoretical,liu2008monte}.
More precisely, we  first decompose $\btheta=(\theta_1,\ldots,\theta_p)$, and construct a sequence of (unnormalized) auxiliary distributions $\eta_1(\theta_1), \eta_2(\theta_1,\theta_2),\ldots,\eta_p(\btheta)$ sequentially approaching the target distribution, i.e., $\eta_p(\btheta) = \gamma(\btheta)$.
We then initialize by drawing $n$ samples, $\theta_1^{(1)}, \cdots, \theta_1^{(n)}$, from a (normalized) proposal distribution $q_1(\theta_1)$
and attach to each with weight $w_1^{(i)} = \eta_1(\theta_1^{(i)})/q_1(\theta_1^{(i)})$. The average weight $\overline{w}_1$
serves as an estimate of the normalizing constant of $\eta_1(\theta_1)$. 
In the next step, we can either resample the obtained ``particles" $\{(\theta_1^{(i)}, w_1^{(i)})\}_{i = 1}^n$ with probability proportional to, say $(w_1^{(i)})^\alpha$ with $\alpha \in [0, 1]$, and modify the new weights to $(w_1^{(i)})^{1-\alpha}$, or proceed directly to sample $\theta_2^{(i)}$ from a (normalized) user-chosen sampling distribution $q_2(\theta_2 \mid \theta_1^{(i)})$. We then update the weights
\[w_2^{(i)} = w_1^{(i)} \times \frac{\eta_2(\theta_1^{(i)},\theta_2^{(i)})}
{\eta_1(\theta_1^{(i)})q_2(\theta_2^{(i)} \mid \theta_1^{(i)})} .\]
Similarly, $\overline{w}_2$
is an estimate of the normalizing constant of $\eta_2(\theta_1, \theta_2)$. This sequential update is carried out up to step  $p$, and the final average weight $\overline{w}_p$ is an estimate of the normalizing constant $Z_\gamma$.
It is easy to show that  $\overline{w}_p$ is unbiased, if no resampling is involved, and is always consistent \citep{del2004feynman}. 

We can generalize the above SMC framework to cases where no dimensional changes are involved, that is,  $\eta_1,\ldots,\eta_p$ are all defined on the full space of $\btheta$. In this case SMC looks very similar to path sampling and the parallel WL method. Recent work has demonstrated its potential in normalizing constant estimation for  applications in Bayesian phylogenetics \citep{wang2020annealed}, nonlinear ordinary differential equation (ODE) models, and positron emission tomography (PET) compartmental models \citep{zhou2016toward}. It is generally nontrivial to design an appropriate sequence of auxiliary and sampling distributions.  Various proposals have been documented in the literature.
(i) The geometric path, $\eta_p(\btheta) = \gamma(\btheta)^{\lambda_p}q(\btheta)^{1 - \lambda_p}$ with $0 = \lambda_0 < \lambda_1 < \cdots < \lambda_p = 1$.
(ii) The posterior distribution with partial data \citep{chopin2002sequential}. In this case, a common choice for the sampling distributions is some form of prior/posterior predictive distributions.
(iii) The path of level sets defined by the likelihood function \citep{salomone2018unbiased} or specific functions associated with tasks of rare event estimation \citep{cerou2012sequential}. We note that the selection of the auxiliary distributions in the aforementioned forms can be potentially made automatic using the conditional effective sample size criterion proposed in \citet{zhou2016toward}.

Another related method is the stepping stone (SS) method \citep{xie2011improving}, which has been successfully applied in Bayesian phylogenetics. The stepping stone method takes the prior as the proposal distribution, and employs a geometric sequence of auxiliary distributions. By expanding the ratio $Z_\gamma/Z_q$ into a series of telescopic product, i.e., $Z_r/Z_q = \prod_{j = 1}^p Z_j/Z_{j - 1}$, in which $Z_j$ is the normalizing constant of the auxiliary distribution $\eta_j(\btheta)$, it uses importance sampling to estimate each ratio $r_j = Z_j/Z_{j - 1}$ by
\begin{equation}
\widehat{r}_j = 
\frac{1}{n}\sum_{i = 1}^np(\by\mid\btheta_{ji})^{\lambda_j - \lambda_{j - 1}},
\end{equation}
in which $p(\by\mid\btheta)$ is the likelihood function, and $\btheta_{ji}$ is the $i$-th MCMC samples from $\eta_j(\btheta)$. A generalized stepping stone method is proposed in \citet{fan2011choosing}, which suggests to use a different proposal better tracking the posterior distribution if the prior is too diffuse.
An illustration of SMC, the parallel WL method and the stepping stone method on the Log-Gaussian Cox process is given in Section \ref{subsec:log-gaussian-cox}.

\subsection{Bridge sampling and path sampling}
Bridge sampling provides an efficient way of utilizing samples from both the proposal and target distributions. Given the (unnormalized) target $\gamma(\btheta)$ and  proposal $q(\btheta)$, bridge sampling inserts a bridge distribution $\gamma_{1/2}(\btheta)$ between $\gamma(\btheta)$ and $q(\btheta)$, and estimates the ratio $Z_\gamma/Z_q$ based on the following identity:
\begin{equation}
r = \frac{Z_\gamma}{Z_q} = \frac{\mathbbm{E}_q\left[\gamma_{1/2}(\btheta)/q(\btheta)\right]}{\mathbbm{E}_\gamma\left[\gamma_{1/2}(\btheta)/\gamma(\btheta)\right]}.
\label{eq:bridge-sampling}
\end{equation}
The corresponding bridge sampling estimator is 
\begin{equation}
\widehat{r} = \frac{(1/n_q)\sum_{i = 0}^{n_q}\gamma_{1/2}(\btheta_{qi})/q(\btheta_{qi})}{(1/n_\gamma)\sum_{i = 0}^{n_\gamma}\gamma_{1/2}(\btheta_{\gamma i})/\gamma(\btheta_{\gamma i})},
\end{equation}
in which $\btheta_{q1},\cdots,\btheta_{q n_q}$ and $\btheta_{\gamma1},\cdots,\btheta_{\gamma n_\gamma}$ are $n_q$ samples and $n_\gamma$ samples from $q^\star(\btheta)$ and $\gamma^\star(\btheta)$, respectively. 

The bridge distribution helps create more connections between the target and the proposal. In addition, since bridge sampling also utilizes samples from the target, it helps resolve the issue of underestimating the normalizing constant as illustrated in Figure \ref{subfig:demo-WL} and discussed in Section \ref{subsec:comparison-IS}. However, the efficiency of bridge sampling is still sensitive to the ``distance" between $q^\star(\btheta)$ and $\gamma^\star(\btheta)$.
For simplicity, let us assume $n_q = n_\gamma = n$, and consider the optimal bridge $\gamma_{\text{opt}}(\btheta) = (q^\star(\btheta)^{-1} +  \gamma^\star(\btheta)^{-1})^{-1}$ that minimizes the asymptotic variance of $\log\widehat{r}$ under the assumption that all the samples are independent draws. The corresponding optimal asymptotic variance is:
\begin{equation}
V_{\text{opt}} = \frac{2}{n}\left[\left(\int\frac{2q^\star(\btheta)\gamma^\star(\btheta)}{q^\star(\btheta) + \gamma^\star(\btheta)}d\btheta\right)^{-1} - 1\right] \geq \frac{2}{n}\left[\left(\int 2\min\{q^\star(\btheta), \gamma^\star(\btheta)\}d\btheta\right)^{-1} - 1\right].
\label{eq:bridge-sampling-variance}
\end{equation}
We see that the lower bound increases if we push the proposal $q^\star(\btheta)$ and the target $\gamma^\star(\btheta)$ further apart. 
In contrast, with the help of global jumping algorithms such as MTM, the WL mixture method is insensitive to this separation issue.

We empirically compare the performance of bridge sampling and the WL mixture method (equiped with MTM) on a 20-dimensional multivariate normal distribution. The target is $N(\bm{0}, I_{20})$, and the surrogate (proposal) is $N(\mu\times\bm{1}_{20}, I_{20})$ with $\mu = 1, 2, 3, 4, 5$. The target has been normalized so that the true log normalizing constant is 0. The fixed jumping direction is $\bm{e} = \pm(\mu\times\bm{1}_{20})$, and we use 8 tries in each multiple-try iteration. For simplicity, we substitute the local MCMC moves around the two mixture components by directly sampling from either the target or the surrogate. We set $n_\gamma = n_q = 5,000$, and run 5,000 iterations for the  WL mixture method. The results are summarized in Table \ref{tab:mvn}.
\begin{table}[h]
\centering
\begin{tabular}{ |c|c|c|c|c|c|}
\hline
Method & $\mu = 1$ & $\mu = 2$ & $\mu = 3$ & $\mu = 4$ & $\mu = 5$\\
\hline
WL & 0.00 (0.05) & 0.01 (0.04) & 0.00 (0.04) & -0.00 (0.04) & 0.01 (0.05)\\
\hline
BS & -0.00 (0.11) & 0.21 (3.15) & -0.35 (5.43) & -1.27 (6.93) & 1.60 (7.90)\\
\hline
\end{tabular}
\caption{Comparisons of the WL mixture method and bridge sampling for estimating the logarithm of the integral of the multivariate normal density, 
which is exactly 0 in all cases. The reported values are empirical means and standard deviations (in the bracket) based on 10 independent runs.}
\label{tab:mvn}
\end{table}

We see that the WL mixture method has robust performances for different $\mu$'s, whereas bridge sampling performs worse as the target  and the proposal distributions become more and more separated. We note that the comparison is not entirely fair because we pre-locate the mode of the target for MTM.
Our point is that the WL mixture method  should be classified as an MCMC-based method, and behaves very differently from bridge sampling and other importance sampling based methods. The performance of the WL mixture method  relies on an efficient strategy to sample from the adaptive mixture distribution $\pi_t^\dagger(\btheta)$ (see Equation \eqref{eq:twisted-working-target-constant-WL}), rather than the amount of overlaps between the target and the surrogate distributions. 

It is conceivable that bridge sampling can overcome the separation between the target and the proposal by creating multiple bridge distributions. \citet{gelman1998simulating} pointed out that when the number of bridge distributions goes to infinity, bridge sampling is equivalent to path sampling. Given a \textit{continuous path} $\{\eta_t(\btheta)\}$ from $q(\btheta)$ to $\gamma(\btheta)$ parameterized by $t \in [0, 1]$ (e.g., the geometric path), path sampling utilizes the identity of thermodynamic integration. That is,
 \begin{equation}\nonumber
 \log\frac{Z_\gamma}{Z_q} = \int_{0}^{1}\mathbbm{E}_t\left[\frac{d}{dt}\log \eta_t(\btheta) \right]dt,
 \end{equation}
in which $\mathbbm{E}_t$ is taken with respect to $\eta_t(\btheta)$. The corresponding path sampling estimator can be constructed using numerical integration over $t$ with samples from $\eta_t(\btheta)$.

\subsection{Chib's method}
The method proposed by \cite{chib1995marginal} is effective for estimating normalizing constants for a class of Bayesian models and has been widely adopted. For any $\btheta^\star$ such that $p(\btheta^\ast \mid \by)>0$, we have
\begin{equation}
\log Z_\gamma =  \log p(\by \mid \btheta^\star) + \log p(\btheta^\star)  - \log \gamma^\star(\btheta^\star\mid\by),
\end{equation}
in which $p(\by \mid \btheta^\star)$ and $p(\btheta^\star)$ are the likelihood function and the prior evaluated at $\btheta^\star$, respectively. Consequently, if we can estimate well the normalized posterior density at $\btheta^\star$, that is, $\gamma^\star(\btheta^\star\mid\by)$, we have an estimate of the normalizing constant $Z_\gamma$. 

\citet{chib1995marginal} showed that this is feasible using  outputs from a Gibbs sampler.
For example, suppose  $\btheta$ can be decomposed into two blocks, $\btheta = (\btheta_1, \btheta_2)$, and we have an efficient Gibbs sampler in hand, 
which iteratively samples from the two conditional distributions $\gamma^\star(\btheta_1\mid\btheta_2, \by)$ and $\gamma^\star(\btheta_2\mid\btheta_1, \by)$. We further assume that we can evaluate these two conditional distributions exactly. With the Gibbs outputs $\{(\btheta_1^{(i)}, \btheta_2^{(i)})\}_{i=1}^n$,
we can estimate the normalized posterior density at $\btheta^\star$ as 
\begin{equation}
\widehat{\gamma}^\star(\btheta^\star\mid\by) = \widehat{\gamma}^\star(\btheta_1^\star\mid\by)\gamma^\star(\btheta_2^\star\mid \btheta_1^\star, \by) = \left[\frac{1}{n}\sum_{i = 1}^n \gamma^\star(\btheta_1^\star\mid\btheta_2^{(i)}, \by)\right]\gamma^\star(\btheta_2^\star\mid \btheta_1^\star, \by),
\end{equation}
which utilizes the identity that $\gamma^\star(\btheta_1\mid \by) = \int\gamma^\star(\btheta_1\mid \btheta_2, \by)\gamma^\star(\btheta_2\mid \by)d\btheta_2$. For a better statistical efficiency, it is recommended to select $\btheta^\star$ close to the posterior mode. The above scheme can be generalized to cases in which $\btheta$ is decomposed into an arbitrary number of blocks, and also cases with missing data \citep{chib1995marginal}.

We see that Chib's method is particularly useful and easy to implement when we have an efficient Gibbs sampler with all the conditional distributions being tractable. In Section \ref{subsec:lasso}, we compare  Chib's method and the WL mixture method on the Bayesian Lasso example, in which we indeed have a closed form Gibbs sampler. The performances of the two methods are comparable. \citet{chib2001marginal} extended the method to settings with intractable conditional densities, but its applicability can still be limited if we encounter other types of MCMC algorithms such as Hamiltonian Monte Carlo (HMC), Metropolis-adjusted Langevin algorithm (MALA), etc. In contrast, a major advantage of the WL mixture method is that it can be built on any type of MCMC samplers, and is reasonably easy to implement.

\section{Illustrations}
\label{sec:illustration}
\subsection{Log-Gaussian Cox process}
\label{subsec:log-gaussian-cox}
We consider estimating the normalizing constant of a Log-Gaussian Cox process on the pine forest data set studied in \cite{penttinen1992marked} and \cite{stoyan1994fractals}.
The data contains the locations of 126 Scots pine saplings on a $10\times10\ \text{m}^2$ square (see Figure 10(a) in \citet{moller1998log}). We first standardize the locations into unit square and then discretize the unit square into an $M \times M$ regular grid. Let $\by = \left(y_m\right)_{m \in \left[M\right]^2}$ denote the number of pine saplings in each grid cell, and let $\blambda = \left(\lambda_m\right)_{m \in \left[M\right]^2}$ denote the latent intensity process. We assume the following model:
\begin{equation}\nonumber
[y_m \mid \lambda_m] \sim \text{Poisson}\left(a\lambda_m\right),
\end{equation}
in which $a = M^{-2}$ is the area of each grid cell. The dimension of $\blambda$ is $M^2$, and in this example, we test out $M = 10,\ 20,\ 30$, thus the dimension of the problem is $100,\ 400,\ 900$, respectively. We transform $\btheta = \log\blambda$ so that all the parameters are defined on $\mathbbm{R}$. We specify a Gaussian process prior with constant mean $\mu_0$ and exponential covariance function given as below on $\btheta = \left(\theta_m\right)_{m \in \left[M\right]^2}$,  
\begin{equation}\nonumber
\Sigma_0\left(m, n\right) = \sigma^2 \exp\left(-\frac{1}{M\beta} \left|m - n\right|\right),\ \ \ \ m,n\in\left[M\right]^2,
\end{equation}
where we follow the same parameters setting in \citet{moller1998log}: $\sigma^2 = 1.91$, $\beta = 1/33$ and $\mu_0 = \log(126) - \sigma^2/2$. The Poisson likelihood is
\begin{equation}\nonumber
L\left(\btheta\mid\by\right) = \prod_{m \in \left[M\right]^2} \exp\left(\theta_my_m - a\exp\left(\theta_m\right)\right),
\end{equation}
thus the unnormalized posterior distribution is $\gamma\left(\btheta\mid\by\right) = N\left(\btheta; \mu_0, \Sigma_0\right)L\left(\btheta\mid\by\right)$. An approximate mode $\widehat{\btheta}$ of $\gamma\left(\btheta\mid\by\right)$ is obtained using the Newton-Raphson method. 

With this example, we compare the performances of two classes of methods: (i) single-step methods including the WL mixture method and importance sampling; (ii) multiple-step methods including an adaptive SMC algorithm, the parallel WL method, and the stepping stone (SS) method proposed in \citep{xie2011improving}. The multiple-step methods insert a sequence of auxiliary distributions between the surrogate (proposal) and the posterior, whereas the single-step methods do not. 

For the single-step methods, we use $N(\bm{\mu}_q, \sigma_q^2I)$ as the surrogate, in which $\bm{\mu}_{q} = \widehat{\btheta}$, and $\sigma_q = 1.0, 1.2, 1.3$ for $M = 10, 20, 30$, respectively, in order to approximately match the marginal posterior standard deviations. For the WL mixture method, we use HMC local moves around the mixture component $\gamma$. The gradient of the log likelihood is
\begin{equation}\nonumber
\nabla \log L\left(\btheta\mid\by\right) = \by - a\exp(\btheta),
\end{equation}
and the HMC kernel contains 10 leapfrog steps with step size 0.25. We run in total $S = 5\times 10^4$ iterations for $M = 10$, and $S = 1 \times 10^5$ iterations for $M = 20,\ 30$, with $b = S/2$. In addition, we test out different thresholds $c$ (from 0.10 to 0.30 incremented by 0.05) used in the flat histogram criterion. 
For importance sampling, we use $1\times10^6$ samples to match the computation time of the WL mixture method (see Table \ref{tab:cox-single-result}).

The algorithmic settings of the mutliple-step methods are described below. 
The SMC algorithm is detailed in the supplementary materials. The proposal distribution is the prior distribution $N\left(\btheta; \mu_0, \Sigma_0\right)$. We use $500$ particles, and run 10 HMC rejuvenation steps (see step 2(g) in Algorithm 1 in the supplementary materials) for each auxiliary distribution to diversify the particles. The conditional effective sample size (CESS) adaptation criterion \citep{zhou2016toward} is set to be $\kappa = 0.9$, and a systematic resampling step is carried out if the normalized effective sample size (ESS) drops below 0.5. On average there are 35, 42 and 45 intermediate steps for $M = 10, 20, 30$, respectively. For both the parallel WL method and the stepping stone method, we employ the same sequence of auxiliary distributions adaptively selected by the SMC algorithm, and run $S = 1.5\times10^3$ iterations using the aforementioned HMC kernel invariant to each auxiliary distribution. 
The computation time for the three methods are comparable (see Table \ref{tab:cox-multiple-result}). 

\begin{figure}[h]
\centering
\includegraphics[width=0.42\columnwidth]{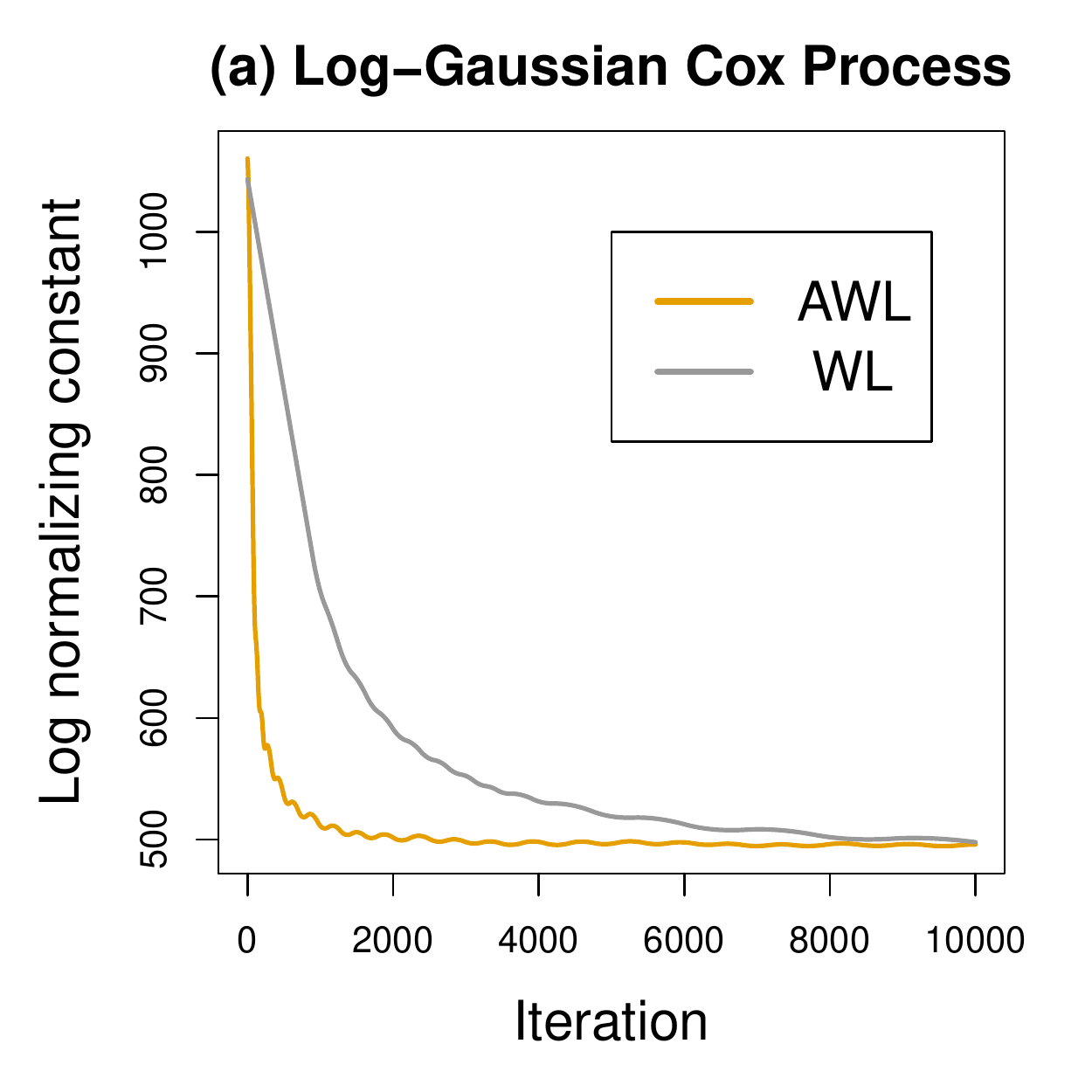}
\includegraphics[width=0.42\columnwidth]{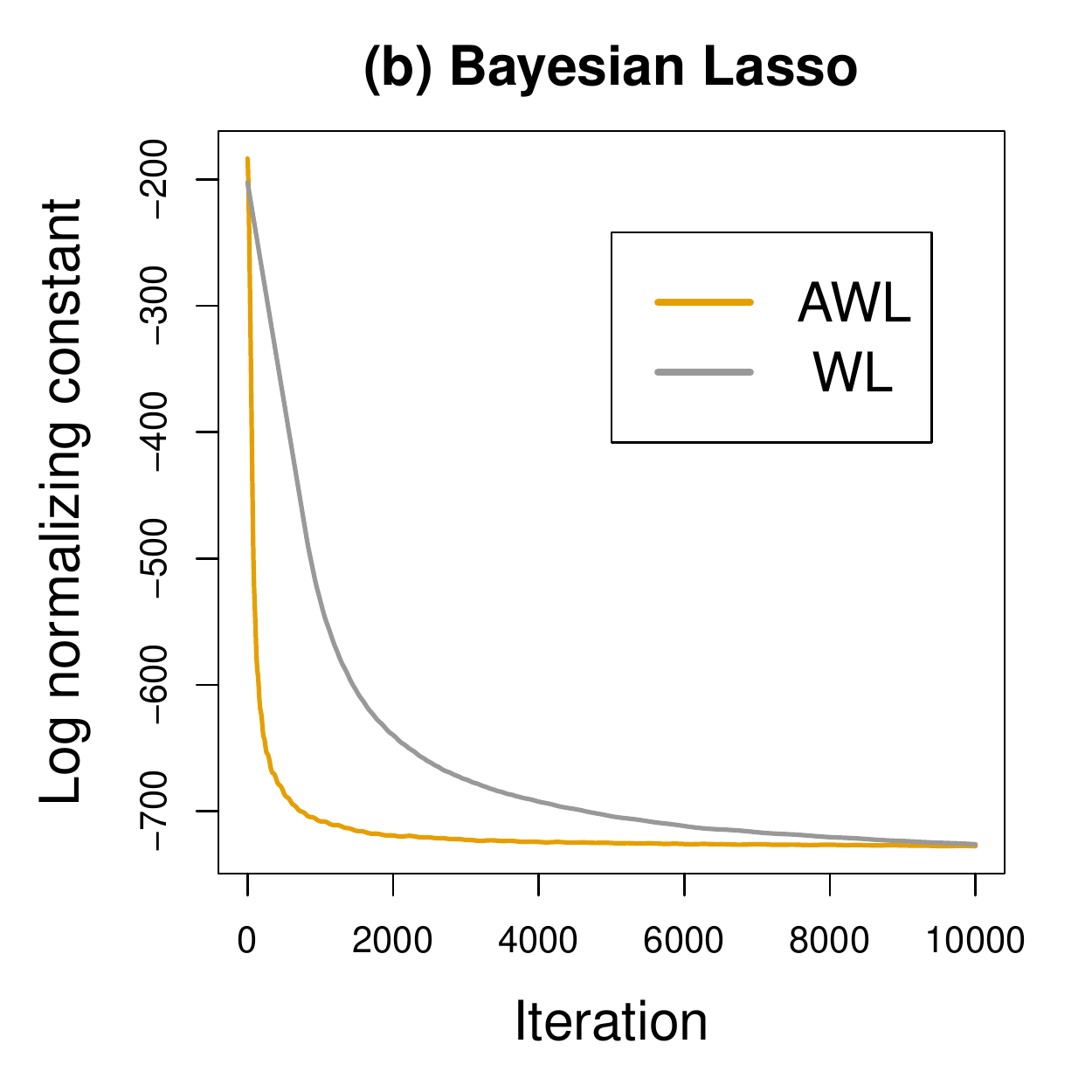}
\caption{Demonstration of the accelerated Wang-Landau algorithm. AWL and WL refer to the accelerated Wang-Landau algorithm and the standard Wang-Landau algorithm, respectively. (a) The Log-Gaussian Cox process discussed in Section~\ref{subsec:log-gaussian-cox}  with $M = 30$. (b) The Bayesian Lasso example discussed in Section~\ref{subsec:lasso} with SNR = 0.1 and $\lambda = 20$.}
\label{fig:AccWL}
\end{figure}

The results are summarized in Tables \ref{tab:cox-single-result} and \ref{tab:cox-multiple-result}. We find that all the three multiple-step methods, as well as the WL mixture method, produced similar estimates of the log normalizing constant across different settings, whereas importance sampling underestimated the log normalizing constant for both $M = 20, 30$, consistent with our discussion in Section \ref{subsec:comparison-IS}. 
The SMC algorithm is the most accurate method for this example, but its implementation is more involved and case-specific than the other two methods. 
The (parallel) WL mixture method also performs reasonably well, having slightly larger standard deviations compared to the SMC algorithm. 
Figure 1 in the supplementary materials shows that the performance of the WL mixture method is robust to the choice of the threshold $c$ used in the flat histogram criterion in the region $[0.1, 0.3]$.
For this  and the Bayesian Lasso example  in the next section, we also compare the convergence speeds of the standard and the accelerated WL algorithms  (see Section \ref{subsec:acceleration}). Figure \ref{fig:AccWL} shows that the accelerated algorithm converges much faster than the standard one. 

\begin{table}[h]
\centering
\begin{tabular}{|c|c|c|c|}
\hline
\multicolumn{4}{|c|}{Log normalizing constant estimates} \\
\hline
Dimension & 100 & 400 & 900 \\
\hline
WL & 474.4 (0.1) & 490.7 (0.3) & 496.6 (0.6) \\
\hline
IS & 474.1 (0.2) & 487.1 (1.0) & 476.0 (1.3) \\
\hline
\hline
\multicolumn{4}{|c|}{Computation time (second)} \\
\hline
Dimension & 100 & 400 & 900 \\
\hline
WL & 15.5 (0.2) & 157.8 (13.4) & 985.4 (60.3) \\
\hline
IS & 17.4 (0.4) & 150.5 (12.6) & 929.8 (55.2) \\
\hline
\end{tabular}
\vspace{0.5cm}
\caption{Log normalizing constant estimates of the Log-Gaussian-Cox process using single-step methods. 
WL and IS refer to the WL mixture method and importance sampling, respectively.
The reported values are empirical means and standard deviations (in the bracket) based on 10 independent runs.}
\label{tab:cox-single-result}
\end{table}

\begin{table}[h]
\centering
\begin{tabular}{|c|c|c|c|}
\hline
\multicolumn{4}{|c|}{Log normalizing constant estimates} \\
\hline
Dimension & 100 & 400 & 900 \\
\hline
SMC & 474.4 (0.1) & 490.7 (0.1) & 497.6 (0.1) \\
\hline
PWL & 474.6 (0.2) & 490.7 (0.3) & 497.3 (0.4) \\
\hline
Stepping stone & 474.6 (0.1) & 491.9 (0.2) & 502.5 (0.8) \\
\hline
\hline
\multicolumn{4}{|c|}{Computation time (second)} \\
\hline
Dimension & 100 & 400 & 900 \\
\hline
SMC & 21.6 (0.7) & 212.2 (17.9) & 1054.2 (66.2) \\
\hline
PWL & 26.9 (0.8) & 189.8 (15.4) & 1123.3 (79.1) \\
\hline
Stepping stone & 20.5 (0.5) & 176.7 (14.6) & 1055.2 (70.4) \\
\hline
\end{tabular}
\vspace{0.5cm}
\caption{Log normalizing constant estimates of the Log-Gaussian-Cox process using multiple-step methods. 
SMC, PWL and Stepping stone refer to the sequential Monte Carlo method, the parallel WL method, and the stepping stone method \citep{xie2011improving}, respectively. The reported time for PWL and the stepping stone method is the computation time without parallelization.
The reported values are empirical means and standard deviations (in the bracket) based on 10 independent runs.}
\label{tab:cox-multiple-result}
\end{table}

\subsection{Hyper-parameter determination for Bayesian Lasso}
\label{subsec:lasso}
We consider the Bayesian Lasso method proposed in \cite{park2008bayesian}, which assumes a hierarchical prior on the linear regression coefficients so that the posterior mode corresponds to the Lasso estimator \citep{tibshirani1996regression}. Given a centered and standardized $n \times p$ design matrix $X$, the response vector $\by$ follows $N(X\bbeta, \sigma^2I_n)$. Following \citet{park2008bayesian}, we specify a prior $N(\bm{0}_p, \sigma^2D_\tau)$ on $\bbeta$, where $D_\tau$ is a diagonal matrix $\text{diag}(\tau_1^2, \cdots, \tau_p^2)$. Besides, we specify an independent hyper-prior $\text{Exp}(\lambda^2/2)$ on each $\tau_j^2$ for $j \in [p]$, and an improper prior $p(\sigma^2)\propto 1/\sigma^2$ on $\sigma^2$.
This completes the full model specification, and the unnormalized posterior distribution is 
\begin{equation}\nonumber
\gamma(\bbeta, \btau, \sigma^2\mid X,\by) = \frac{1}{\sigma^2} N(\by;X\bbeta, \sigma^2I_n)\prod_{j = 1}^p\text{Exp}\left(\tau^2_j\mid\lambda^2/2\right).
\end{equation}
We transform the parameters $\eta_j = \log\tau_j^2$ for $j \in [p]$ and $\xi = \log\sigma^2$ so that all the parameters are defined on $\mathbbm{R}$. 

In this example, we simulate the data set as in \citet{yang2016computational}. Let
\begin{equation}\nonumber
\bbeta^\star = \text{SNR}\sqrt{\sigma_0^2 \frac{\log p}{n}}(2, -3, 2, 2, -3, 3, -2, 3, -2, 3, 0, \cdots, 0)^\intercal \in \mathbbm{R}^p
\end{equation}
with $p = 100$, $n = 500$, $\text{SNR}\in\{0.1, 1, 3\}$ (signal-to-noise ratio), and $\sigma_0^2 = 1$. The dimension of the posterior distribution is $2\times p + 1 = 201$. The design matrix $X$ is generated from a centered multivariate normal distribution with covariance matrix $\Sigma_{ij} = \exp(-|i - j|)$. The response variable $\by$ is generated from $N(X\bbeta^\star, \sigma_0^2I_n)$. The task is to estimate the marginal likelihood of data for a set of regularization parameters $\lambda \in \{5, 10, 15, 20\}$ under different SNRs. 

\begin{figure*}
\centering
\includegraphics[width=0.21\columnwidth]{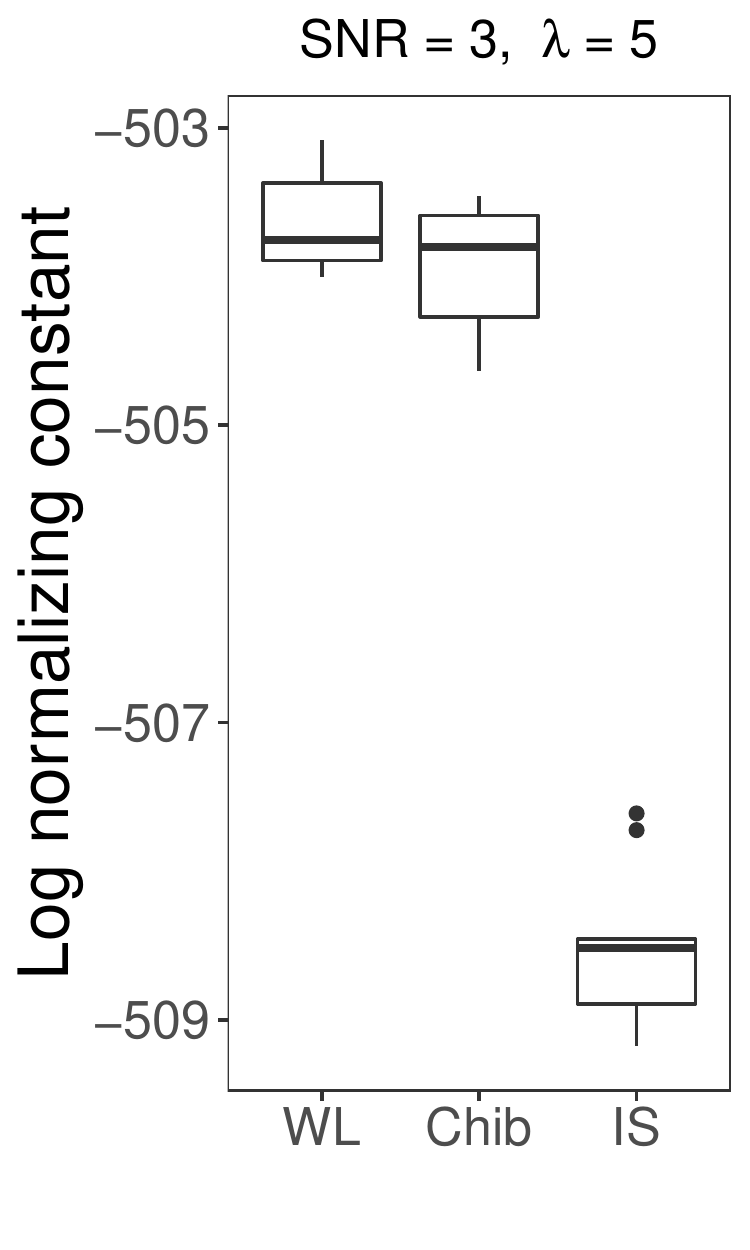}
\includegraphics[width=0.21\columnwidth]{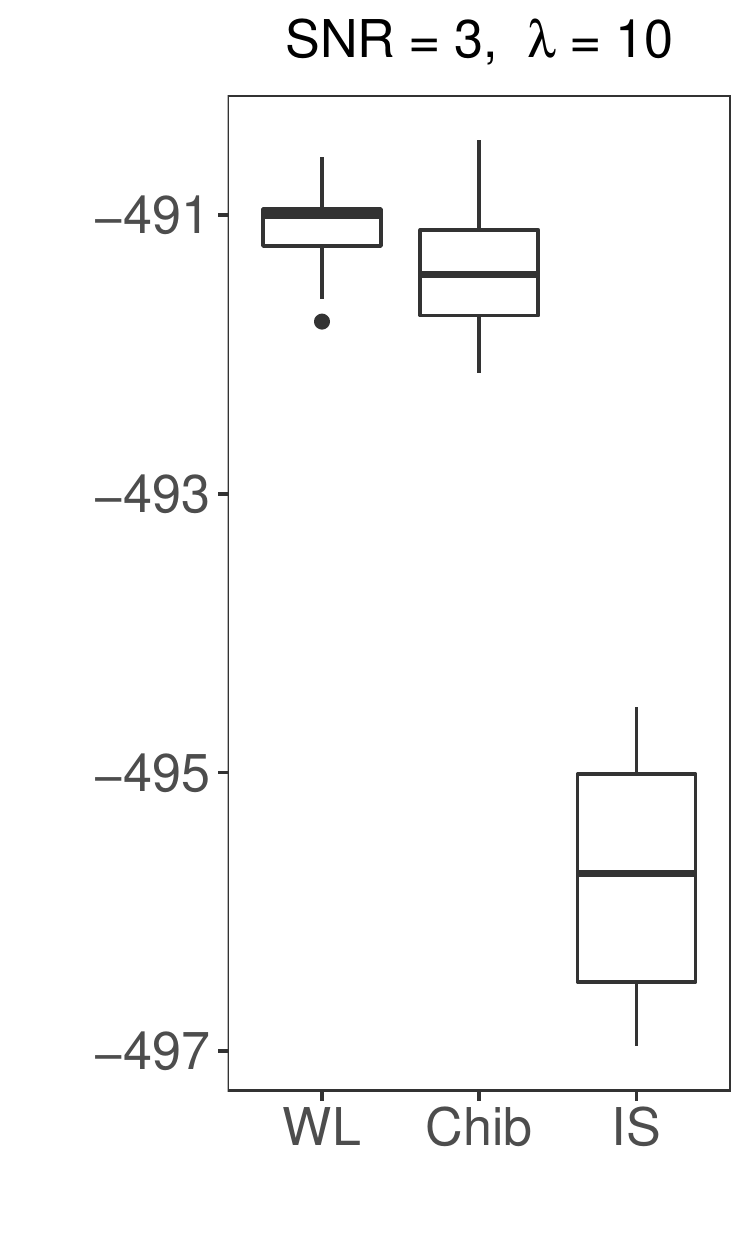}
\includegraphics[width=0.21\columnwidth]{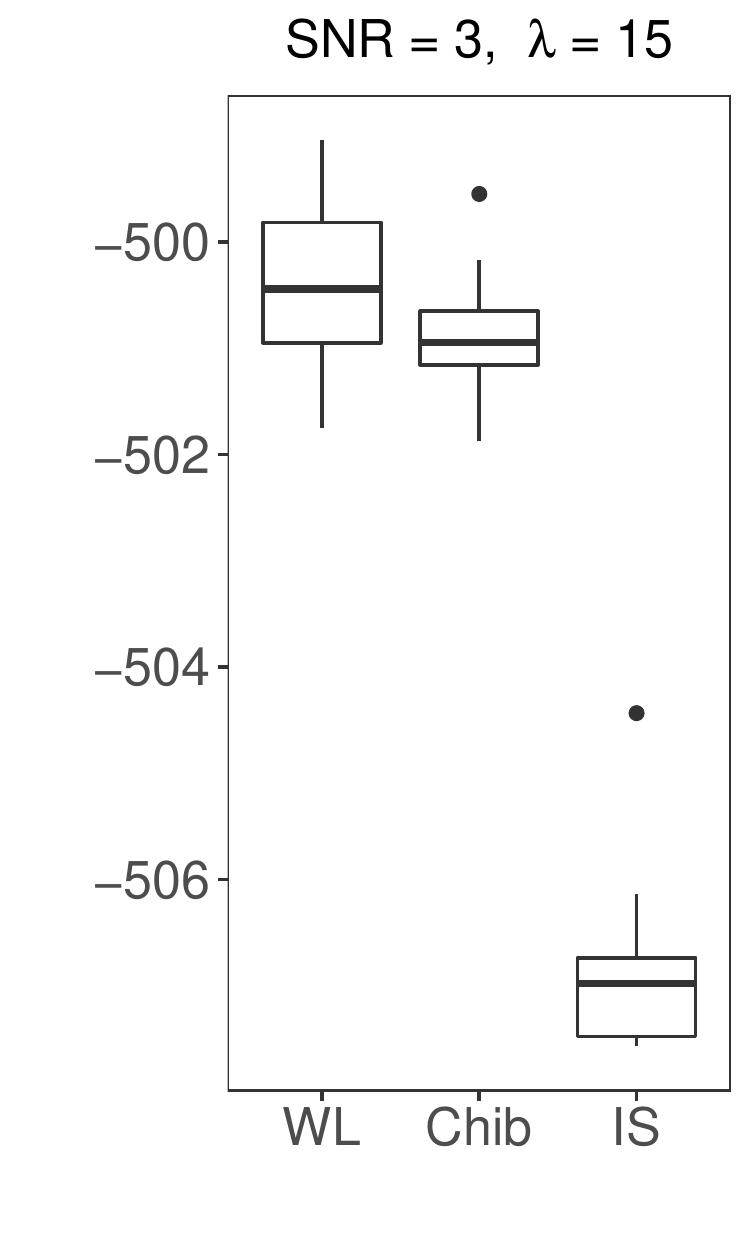}
\includegraphics[width=0.21\columnwidth]{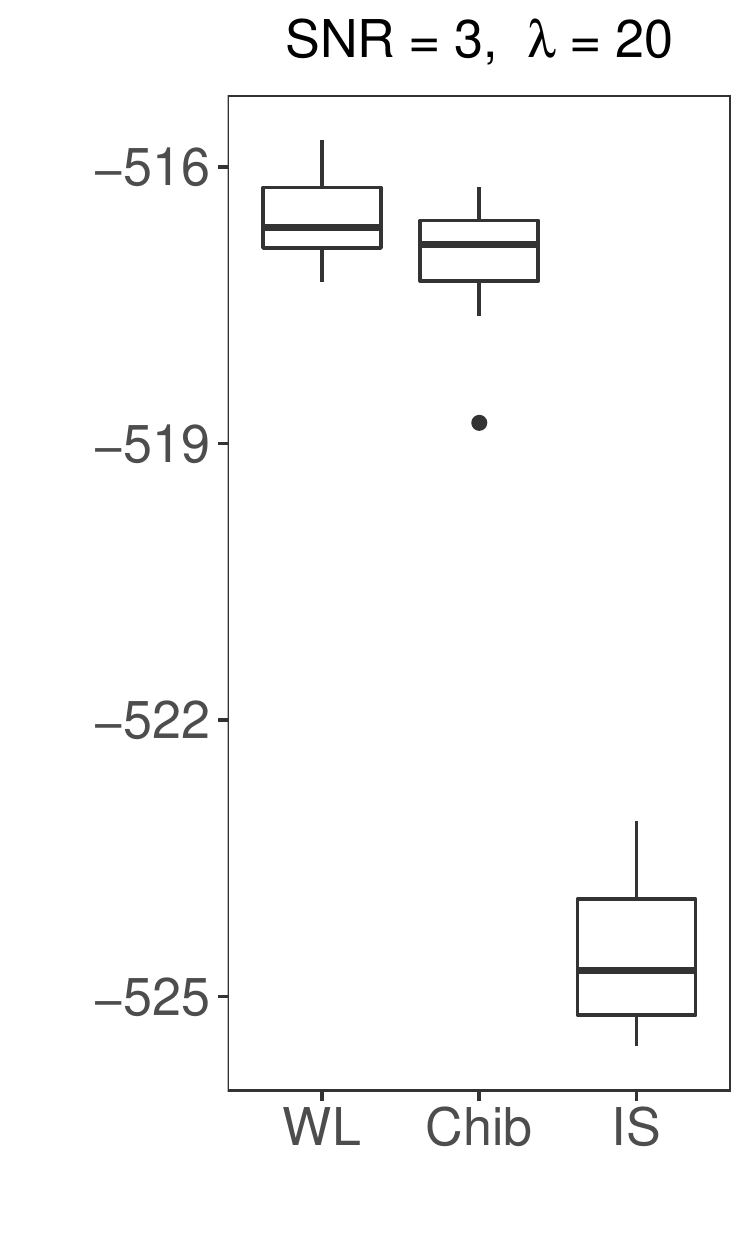}
\includegraphics[width=0.21\columnwidth]{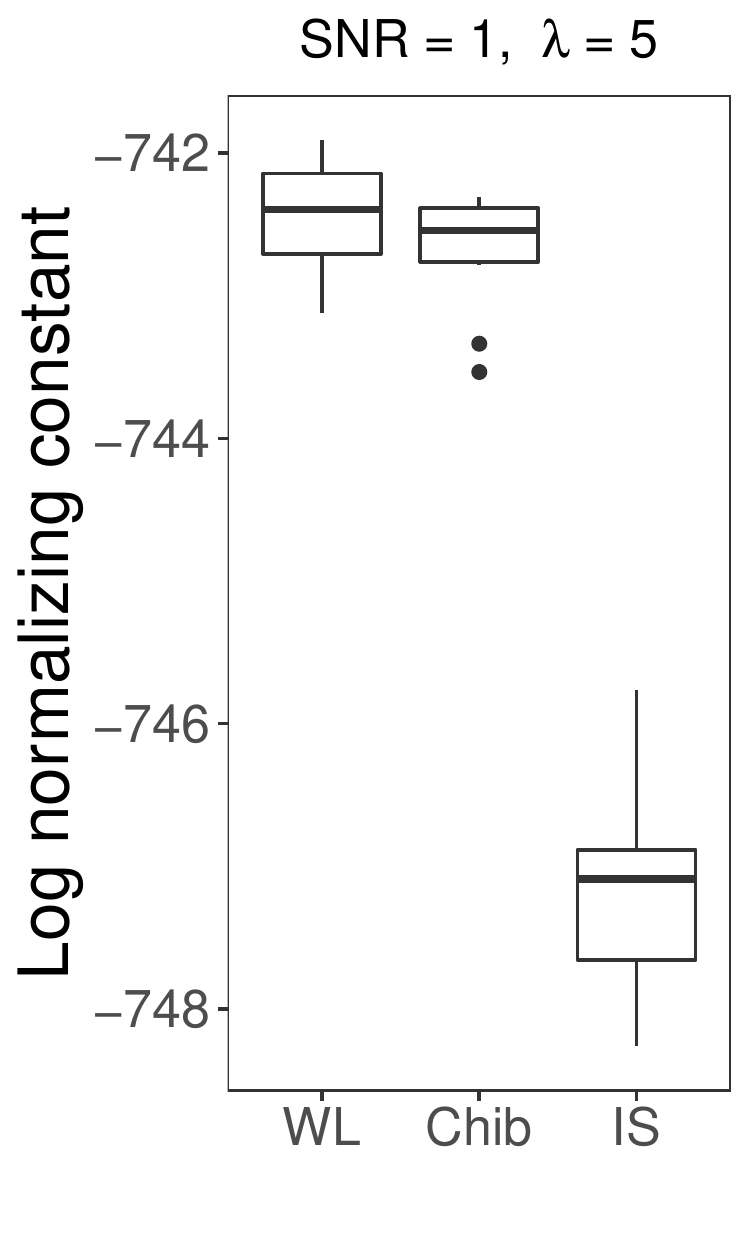}
\includegraphics[width=0.21\columnwidth]{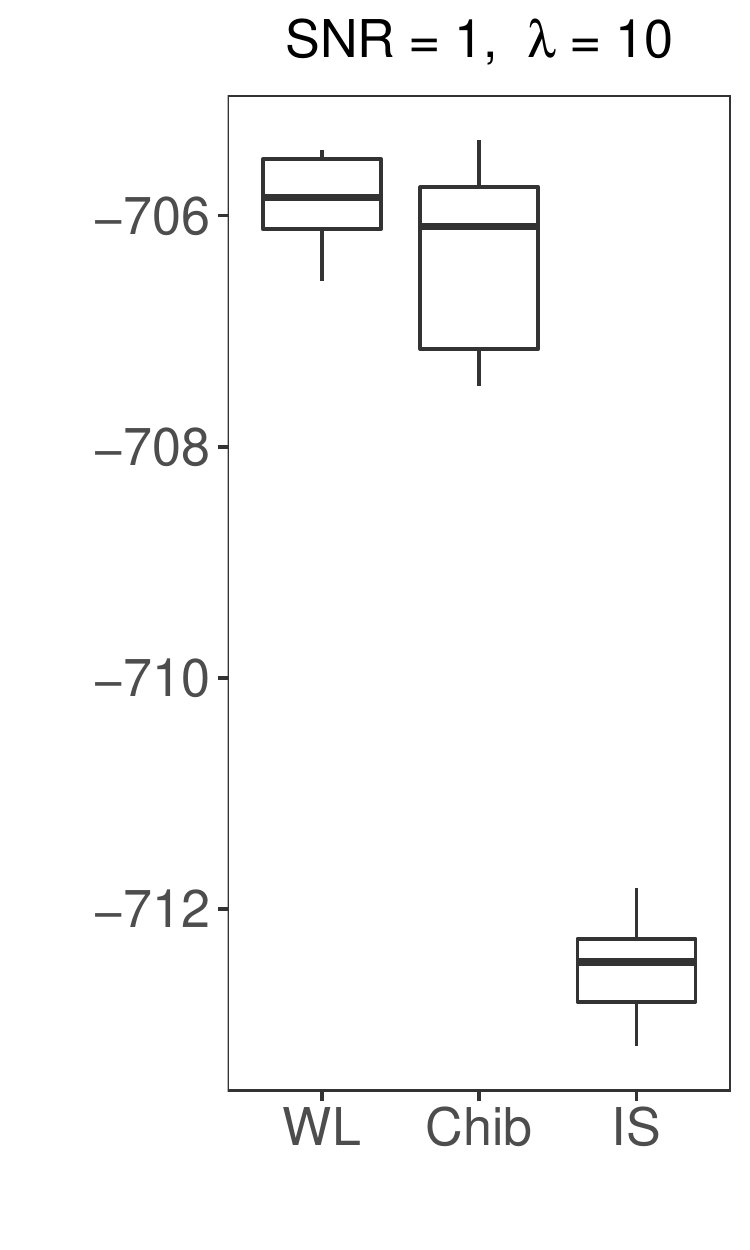}
\includegraphics[width=0.21\columnwidth]{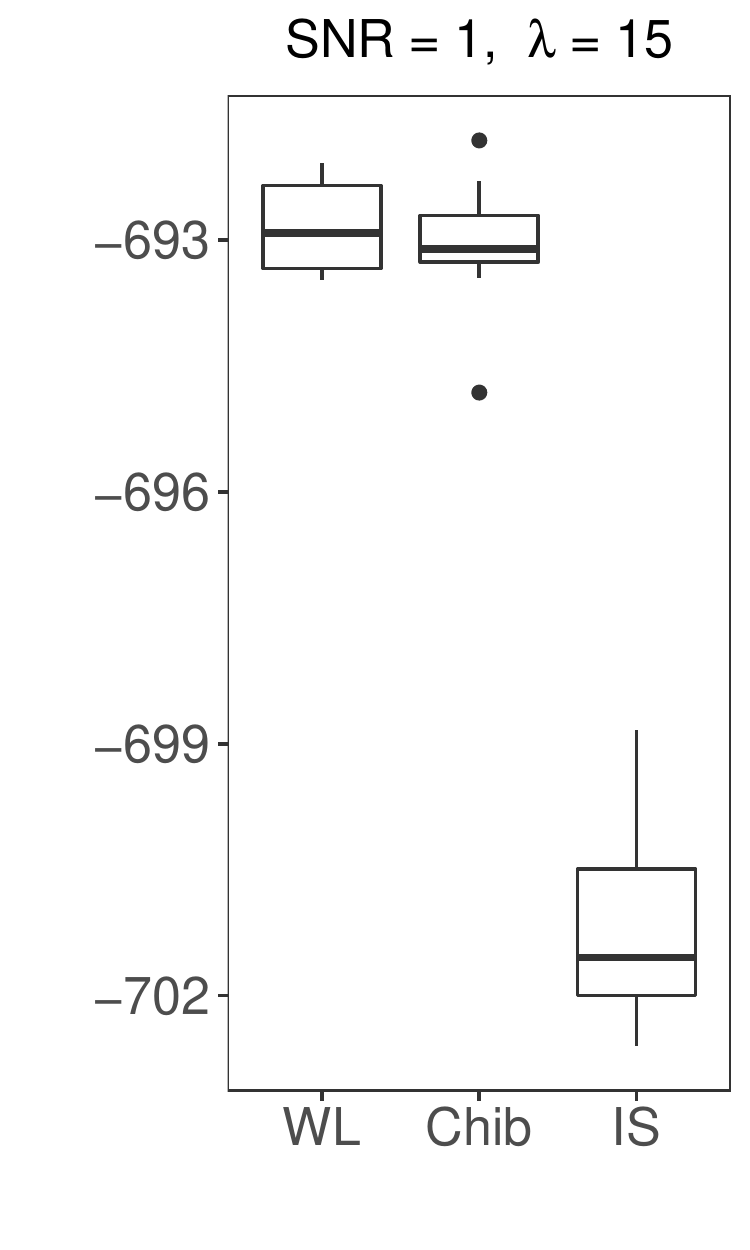}
\includegraphics[width=0.21\columnwidth]{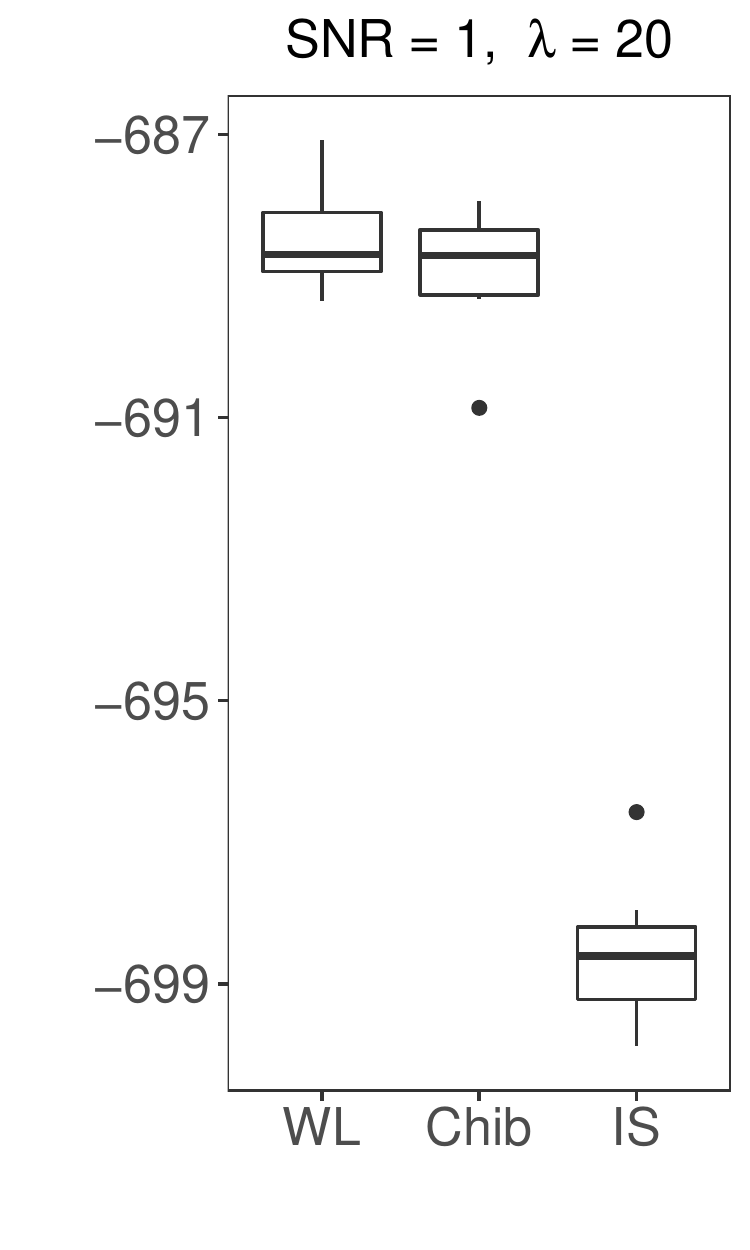}
\includegraphics[width=0.21\columnwidth]{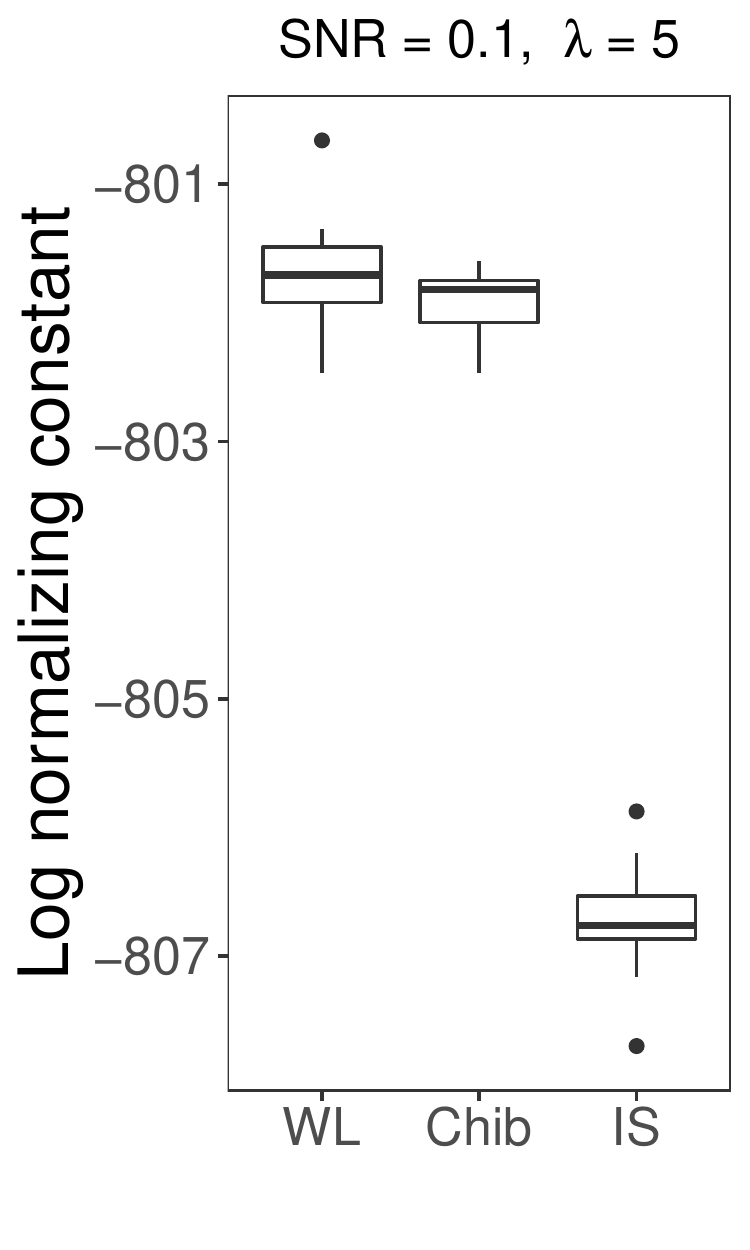}
\includegraphics[width=0.21\columnwidth]{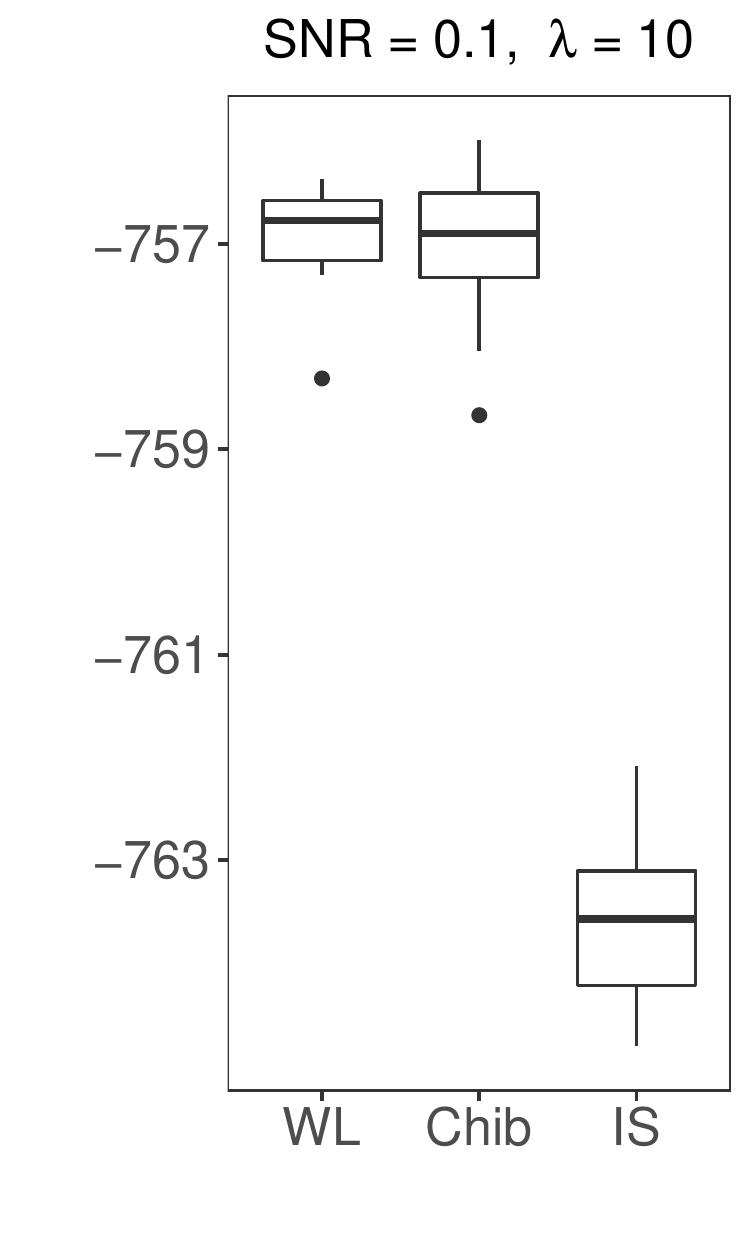}
\includegraphics[width=0.21\columnwidth]{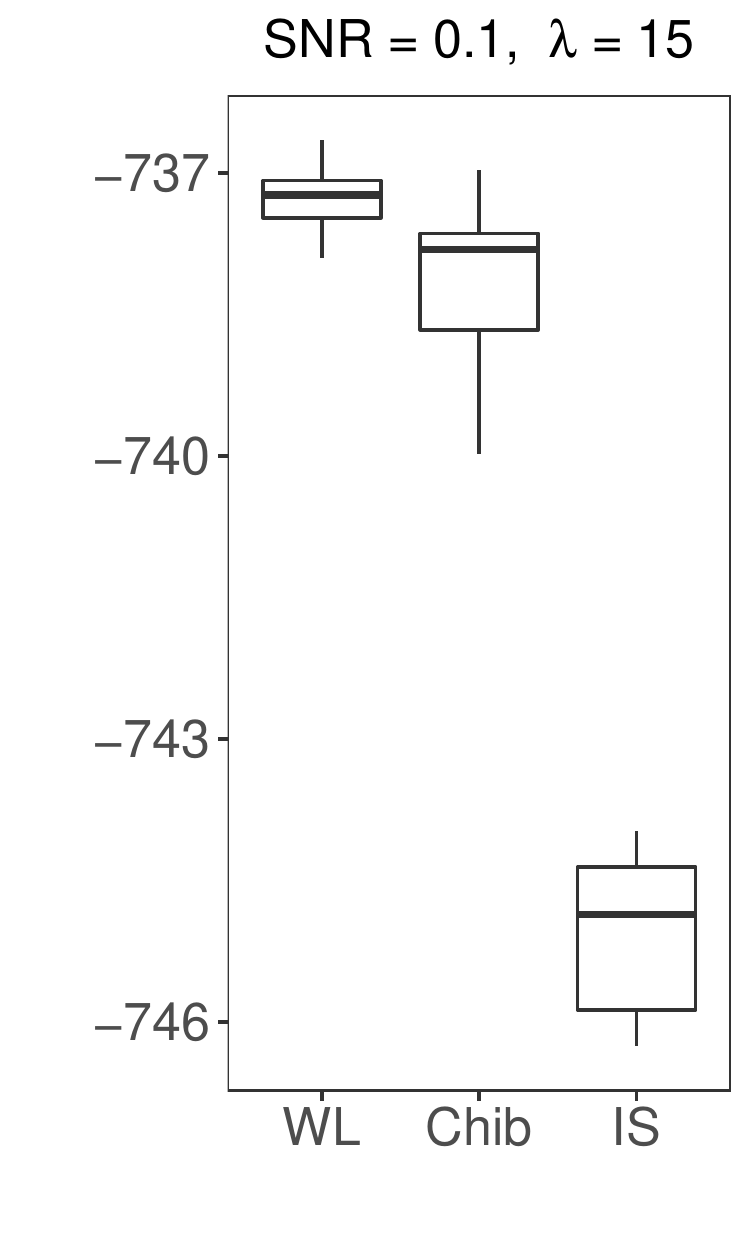}
\includegraphics[width=0.21\columnwidth]{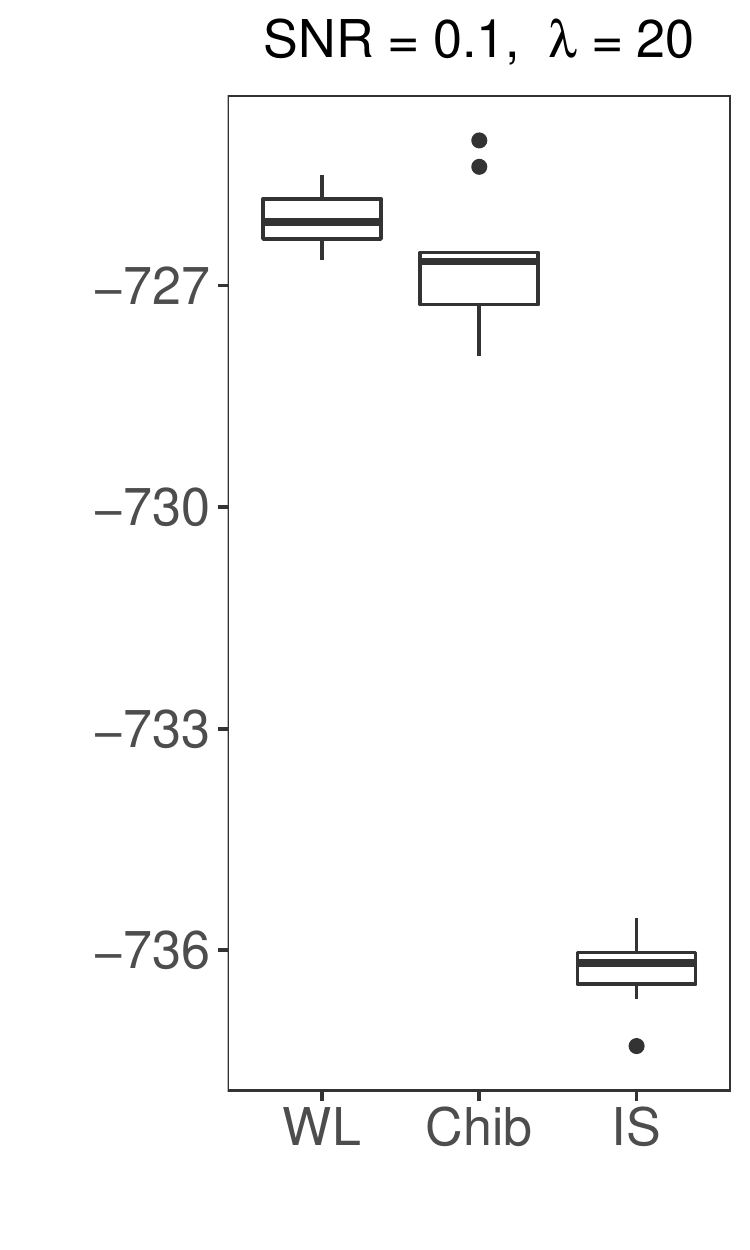}
\caption{Results summary of the Bayesian Lasso example. WL, Chib and IS refer to the WL mixture method, Chib's method and importance sampling, respectively. The box plots are based on 10 independent runs of the algorithms. The computation time for the WL mixture method, Chib's method, and importance sampling are 32.1 ($\pm0.6$) seconds, 43.5 ($\pm2.0$) seconds and 47.1 ($\pm1.2$) seconds, respectively.}
\label{fig:Bayesian-Lasso-result}
\end{figure*}

We compare the WL mixture method, Chib's method and importance sampling for this example. 
The surrogate distribution used in the WL mixture method, which is also the proposal distribution used in importance sampling, is constructed using the variational approximation discussed in Section \ref{subsec:VA}. We consider the Normal mean-field variational family where $q(\beta_j)$ is $N(m_j, s_j^2)$, $q(\eta_j)$ is $N(\phi_j, \zeta_j^2)$, and $q(\xi)$ is $N(u, v^2)$. The CAVI updates are summarized in the supplementary materials.
For the WL mixture method, the Gibbs move proposed in \cite{park2008bayesian} is used to move around the mixture component $\gamma$. For completeness, we detail  below
the  conditional posterior distributions required by the Gibbs sampler:
\begin{equation}\nonumber
\begin{aligned}
&\hspace{-0.07cm}\left[\bbeta\mid\text{rest}\right]  \sim N(C_\tau^{-1}X^\intercal \by, \sigma^2C_\tau^{-1}), \ \ \ C_\tau = X^\intercal X + D_\tau^{-1},\\
&[\tau_j^{-2}\mid\text{rest}]  \sim \text{Inverse-Gaussian}(\lambda\sigma/|\beta_j|, \lambda^2),\\
&[\sigma^2\mid\text{rest}] \sim \text{Inv-}\chi^2\left(n + p, {\left(||\by- X\bbeta||_2^2 + \bbeta^\intercal D_\tau^{-1}\bbeta\right)}/{(n + p)}\right).
\end{aligned}
\end{equation}
We run a total of $S = 1\times10^4$ iterations, and set $b = 2,000$. For importance sampling, we use $5\times10^5$ samples. For Chib's method, the parameters are partitioned into three blocks, $\bbeta$, $\btau^2$ and $\sigma^2$, and we run the Gibbs sampler for $5\times10^3$ iterations and treat the first 10\% samples as burn-in. The computation time for the three methods are comparable (see the caption in Figure \ref{fig:Bayesian-Lasso-result}),

The results are summarized in Figure \ref{fig:Bayesian-Lasso-result}. We see that under all settings, the WL mixture method and Chib's method produced similar and stable estimates of the log normalizing constant, whereas importance sampling underestimated the log normalizing constant. 
The regularization parameter that maximizes the marginal likelihood of data are $\lambda = 10, 20, 20$ for $\text{SNR} = 3, 1, 0.1$, respectively, which corresponds to our intuition that it requires more regularization for estimating the regression coefficients when there exists larger noises in the data.  

\subsection{Logistic regression}
We consider a Bayesian logistic regression model for the classic German credit data set (available
from the UCI repository \citep{frank2011uci}). There are in total $n = 1,000$ personal records in the data set. For each records, there are 24 associated attributes including sex, age, and credit amount. The binary response variable $\by$ indicates good or bad credit risks. Let $X_{n\times p}$ be the design matrix after we standardize all the predictors. In particular, we include an intercept and all pairwise interactions. The dimension of the problem is $p = 24 + 24\times 23/2 + 1 = 301$. We consider the following logistic regression model:
\begin{equation}
\label{eq:logistic-prob}
\mathbbm{P}(y_i = 1\mid\bx_i,\alpha,\bbeta) = \frac{\exp\left(\alpha + \bbeta^\intercal \bx_i\right)}{1 + \exp\left(\alpha + \bbeta^\intercal \bx_i\right)},
\end{equation}
in which $y_i \in \{0, 1\}$, $\bx_i \in \mathbbm{R}^{300}$, $\alpha \in \mathbbm{R}$, $\bbeta\in \mathbbm{R}^{300}$, $i \in [n]$. All the observations are assumed to be independent. We set up similar priors on the parameters as in \citet{heng2019unbiased},
\begin{equation}\nonumber
[\alpha\mid s^2] \sim N\left(0, s^2\right),\ \ \ [\bbeta\mid s^2]\sim N\left(\bm{0}_{300}, s^2I_{300}\right), \ \ \ s^2\sim \text{Exp}(\lambda),
\end{equation}
with $\lambda \in \{0.01, 1.00\}$. This
leads to the unnormalized posterior distribution:
\begin{equation}\nonumber
\begin{aligned}
\gamma(\alpha, \bbeta, s^2\mid\by, X) & = p(\alpha, \bbeta\mid s^2)p(s^2)\prod_{i = 1}^np(y_i\mid\bx_i)\\
&  = \lambda e^{-\lambda s^2}N(\alpha;0, s^2)\prod_{j = 1}^{300}N(\bbeta_j;0, s^2)\prod_{i = 1}^n
p_i^{y_i}(1 - p_i)^{1 - y_i},
\end{aligned}
\end{equation}
in which $p_i = \mathbbm{P}(y_i = 1 \mid\bm{x}_i, \alpha, \bbeta)$ as defined in Equation \eqref{eq:logistic-prob}. We transform $s^2$ to the logarithmic scale $\log s^2$ so that all the parameters are defined on $\mathbbm{R}$. The task is to estimate the log normalizing constant of $\gamma$. 

We compare the WL mixture method and bridging sampling (BS). We use the same surrogate (proposal) distribution, constructed by the Laplace approximation method detailed below, for both methods. 
We first run an HMC algorithm to obtain posterior samples from $\gamma(\alpha, \bbeta,  \log s^2\mid\by, X)$. Then, we fit a multivariate normal distribution on the posterior samples, and choose it as the surrogate distribution. Each HMC step contains 10 leapfrog steps with step size adjusted to be 0.03. For bridge sampling, we use the R package \textit{bridgesampling} \citep{gronau2017bridgesampling}, and obtain $n$ samples from the posterior using RStan \citep{rstan}. Correspondingly, we run $2\times n$ iterations for the WL mixture method so that approximately we also use $n$ samples from the posterior.

\begin{figure}[h]
\centering
\includegraphics[width=0.45\columnwidth]{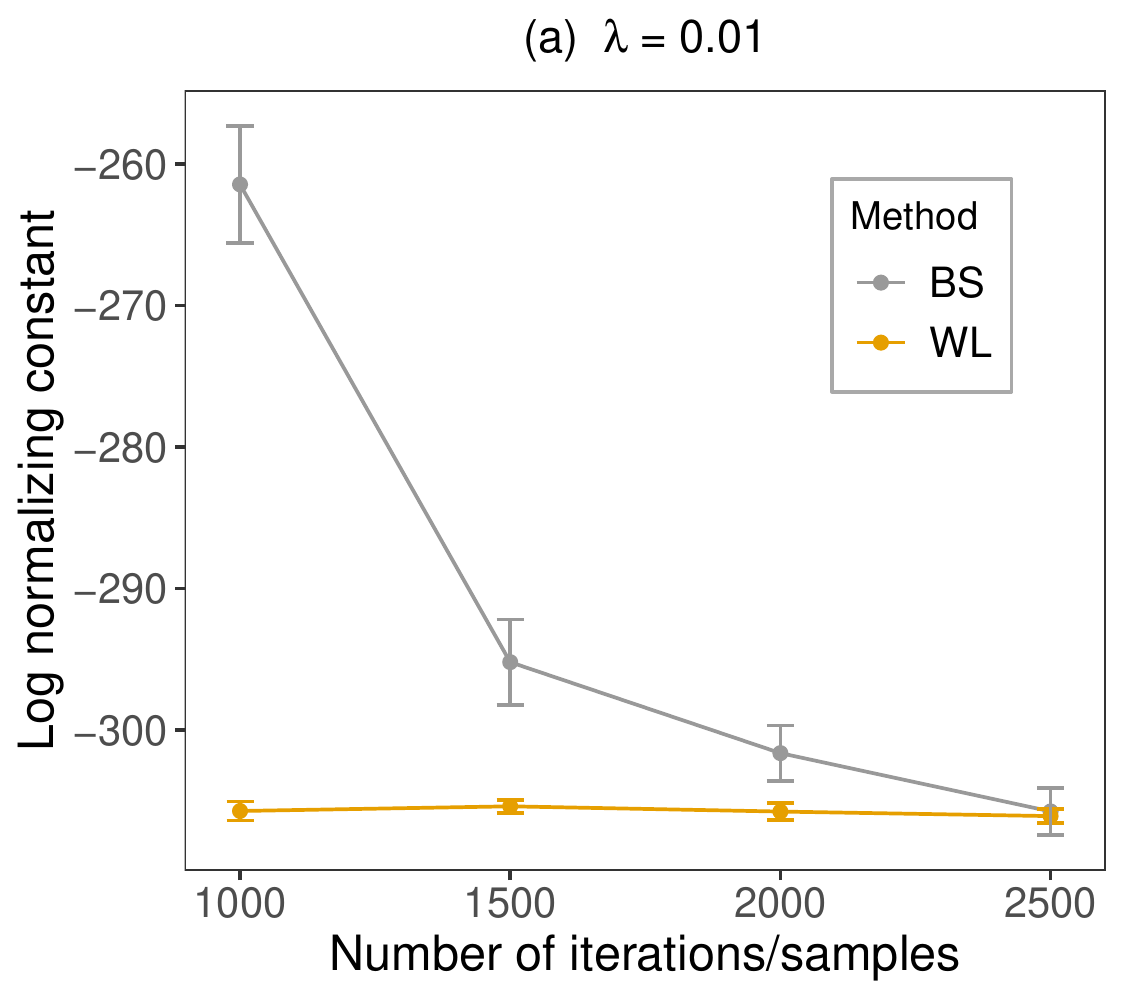}
\includegraphics[width=0.45\columnwidth]{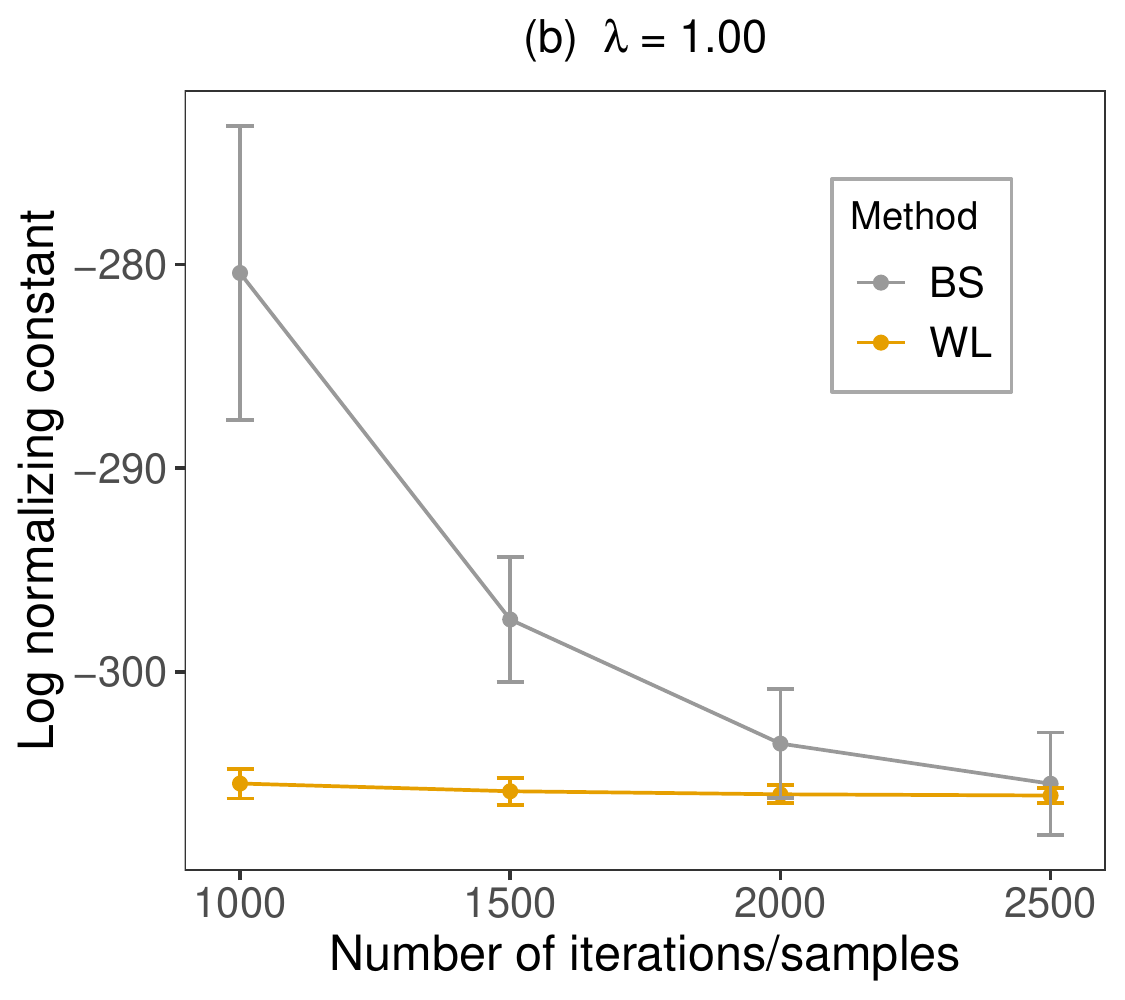}
\caption{Comparison between the  WL mixture method and bridge sampling (BS). For bridge sampling, the $x$-axis represents the number of samples we draw from the posterior and the proposal distributions. For the WL mixture method, the $x$-axis represents half of the total number of iterations we run (see the second to last paragraph in this section). The error bars represent the standard deviations of the log normalizing constant estimates based on 10 independent runs.}
\label{fig:german-credit-result}
\end{figure}

For this example, we test out $n = 1000, 1500, 2000, 2500$ for $\lambda \in \{0.01, 1.00\}$. The results are summarized in Figure \ref{fig:german-credit-result}. We see that the  WL mixture method has a much better estimation efficiency compared to bridge sampling. Bridge sampling approaches to the vicinity of the correct estimate only after 2,500 iterations/samples for both cases $\lambda = 0.01$ and $\lambda = 1.00$. 

\subsection{g-Prior variable selection}
We compare the performance of MTM-RJMCMC proposed in Section \ref{subsec:multiple-try-RJMCMC} and that of a standard birth-and-death RJMCMC (BD-RJMCMC, detailed below) in the setting of Bayesian variable selection for the pollution data set \citep{mcdonald1973instabilities}. The response variable $\by$ is the age-adjusted mortality rate obtained for the years 1959-1961 in 201 standard metropolitan statistical areas. There are in total $n = 60$ observations. The design matrix $X$ contains $p = 15$ predictors including the average annual precipitation, the average temperature in January and July, and the population per household. We consider the standard linear model assuming that $[\by\mid X, \bbeta, \sigma^2]$ follows $N\left(X\bbeta, \sigma^2I\right)$.
We center the response variable $\by$ so that there is no intercept in the model, and standardize each predictor in the design matrix $X$. 

Let $\bgamma \in \{0,1\}^{p}$ be the binary indicator such that $\gamma_j = 1$ represents that the predictor $X_j$ is selected into the model.
We employ the g-prior on $\bbeta$:
\begin{equation}\nonumber
\left[\bbeta_{\bgamma} \mid \bgamma, \sigma^2 \right] \sim N\left(\bm{0}_{\bgamma}, g\sigma^2\left(X_{\bgamma}^\intercal X_{\bgamma}\right)^{-1}\right).  
\end{equation}
The g-prior enables us to integrate out $\bbeta$ so that we can obtain the marginal distribution of $\bgamma$:
\begin{equation}
\label{eq:g-prior-marginal-gamma}
p(\bgamma\mid \by, X) \propto (g + 1)^{-q_{\bgamma}/2}\left[\by^\intercal \by - \frac{g}{g + 1}\by^\intercal X_{\bgamma}\left(X_{\bgamma}^\intercal X_{\bgamma}\right)^{-1}X_{\bgamma} \by\right]^{-n/2},
\end{equation}
in which $q_{\bgamma}$ denotes the number of selected predictors. We see that $g$ controls the sparsity of the model, and a larger $g$ induces a sparser model. For $\sigma^2$, we use a noninformative prior $p\left(\sigma^2\right) \propto 1/\sigma^2$. This completes the full model specification. The task is to estimate the marginal probability of each predictor being selected. The ground truth is obtained by enumerating all 32,768 possible $\bgamma$ and calculating the marginal probability using Equation \eqref{eq:g-prior-marginal-gamma}. To compare MTM-RJMCMC and BD-RJMCMC, we pretend that we do not have the privilege to integrate out $\bbeta$, thus we will sample from the trans-dimensional joint posterior distribution $p\left(\bbeta_{\bgamma}, \bgamma, \sigma^2\mid \by, X\right)$.

We use the Gibbs sampler to iterate between the following conditional distributions:
\begin{equation}\nonumber
\begin{aligned}
\left[\bbeta_{\bgamma}, \bgamma\mid \sigma^2, \by, X\right] & \sim \left(2\pi g\sigma^2\right)^{-\frac{q_{\bgamma}}{2}}\left| X_{\bgamma}^\intercal{X_{\bgamma}}\right|^{\frac12}\exp\left(-\frac{1}{2\sigma^2}\left[\frac{g + 1}{g}\left|\left|X_{\bgamma}\bbeta_{\bgamma}\right|\right|^2 - 2\bbeta_{\bgamma}^\intercal X_{\bgamma}^\intercal \by\right]\right),\\
\left[\sigma^2 \mid \by, X, \bbeta_{\bgamma}, \bgamma\right] & \sim \text{Inv-Gamma}\left(\frac{n + q_{\bgamma}}{2}, \frac{1}{2}\left[\frac{1}{g}\left|\left|X_{\bgamma}\bbeta_{\bgamma}\right|\right|^2 + \left|\left|\by - X_{\bgamma} \bbeta_{\bgamma}\right|\right|^2\right]\right).
\end{aligned}
\end{equation}
Given $\bgamma_t$, the jumping rule for $\bgamma_{t + 1}$ as described below is the same for both algorithms. We first flip a coin to decide whether we stay in the current model ($\bgamma_{t + 1} = \bgamma_t$) or move to a different model ($\bgamma_{t + 1} \neq \bgamma_t$). If we choose to leave the current model (a trans-dimensional move), we randomly move into a higher dimension (add a predictor) or move into a lower dimension (exclude a predictor) with equal probability 0.5. When the chain is at the boundary  ($q_{\bgamma}$ is 1 or 15), the proposal going out of the range is automatically rejected.

\begin{figure}[h]
\centering
\includegraphics[width=1.0\columnwidth]{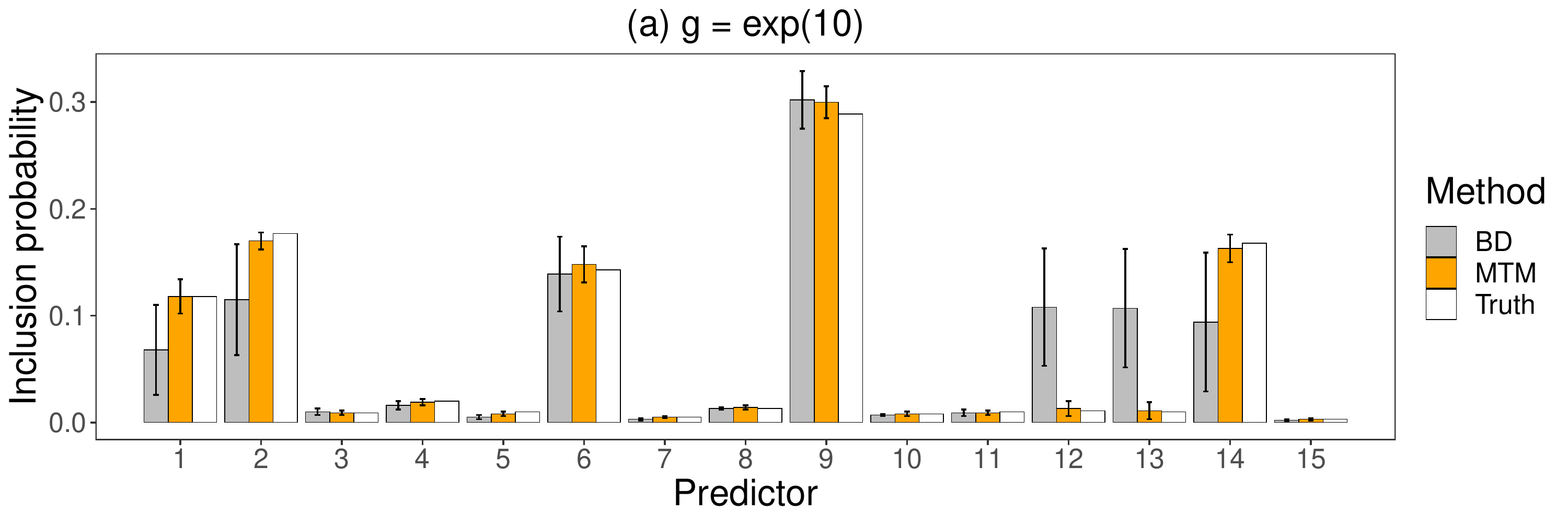}
\includegraphics[width=1.0\columnwidth]{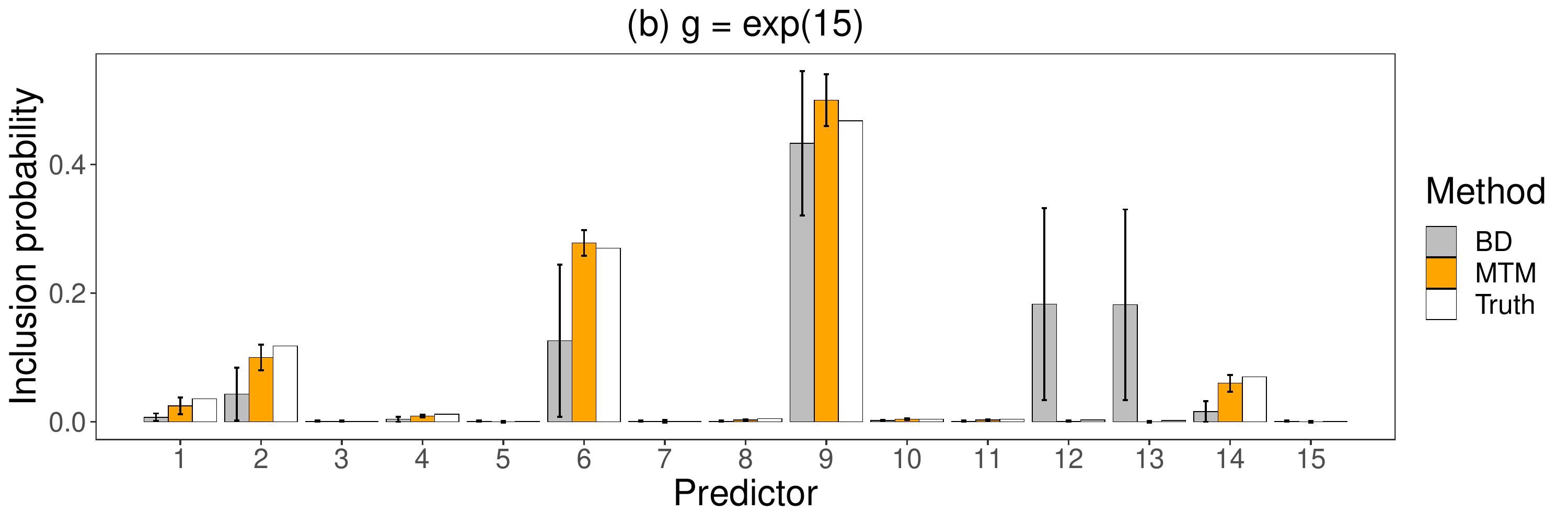}
\caption{Estimates of the marginal probability of each predictor being selected. The bar plots are based on 10 independent runs of both algorithms. The computation time for MTM-RJMCMC and BD-RJMCMC are 19.0 ($\pm1.0$) seconds and 29.8 ($\pm1.2$) seconds, respectively.}
\label{fig:g-prior-variable-prob}
\end{figure}

Given $\bgamma_{t + 1}$,  for the within-dimensional move  ($\bgamma_{t + 1} = \bgamma_t$), we implement an Metropolis-within-Gibbs step, with proposal distribution $N(0, 0.5^2)$, to sequentially update each coordinate of $\bbeta_{\gamma_t}$.
For the trans-dimensional move, MTM-RJMCMC  and BD-RJMCMC use different proposals. For MTM-RJMCMC, we follow the fixed-directional jumping mechanism detailed in Algorithm \ref{alg:mtm-RJMCMC}. Since we only add or remove one predictor in each trans-dimensional move, the algorithm requires only one auxiliary variable. We choose the auxiliary distribution to be $N(0, 1)$. We sample the jumping distance $r$ from $N(1, 1)$, and  set the number of tries to be $m = 5$. For BD-RJMCMC, if we choose to add a predictor, we propose it from $N\left(0, 0.5^2\right)$. We run $5\times 10^4$ iterations for MTM-RJMCMC and $1.5\times 10^5$ iterations for BD-RJMCMC so that the computation time for the two algorithms are comparable (see the caption in Figure \ref{fig:g-prior-variable-prob}). For both algorithms, we burn-in the first 10\% samples.

The estimation results for $g = \exp(10)$ and $g = \exp(15)$ are summarized in Figure \ref{fig:g-prior-variable-prob}. We see that MTM-RJMCMC  produced more accurate estimation results than BD-RJMCMC. In particular, we notice that BD-RJMCMC might have been stuck in a local mode thus mistakenly selected two wrong predictors $X_{12}$ and $X_{13}$. Intuitively, the directional jumping in MTM-RJMCMC is much more informative than the blind proposal used in BD-RJMCMC, thus preventing the algorithm from getting stuck in local modes. 

\section{Concluding Remarks}
\label{sec:conclusion}
We have described a general strategy to construct a mixture of the unnormalized posterior distribution and a surrogate distribution with a known normalizing constant to estimate the model likelihood.
Such a mixture formulation allows us to use the generalized WL algorithm and the MTM machinery for fast MCMC mixing and accurate estimation of the unknown normalizing constant. We have also designed acceleration schemes to further improve its performance.

By efficiently jumping back and forth between the posterior and the surrogate distributions, possibly with the help of mode jumping algorithms such as MTM, the performance of the WL mixture method is less sensitive to the potential separation between the posterior and the surrogate distributions compared to importance sampling based methods. The WL mixture method also has more general applicability compared to Chib's method, when the sampler of the posterior involves more sophisticated MCMC steps beyond the closed-form Gibbs sampler (i.e., all conditional distributions are easy to sample from) or standard Metropolis-Hasting algorithms. In addition, the WL mixture method requires less effort in delicate tuning in its implementation compared to other advanced methods such as  path sampling, reversible jump MCMC, and sequential Monte Carlo methods.

There are several future directions that we would like to follow. First, although we have shown the power of the WL mixture method, a rigorous theoretical framework is required to better understand the nature of the method. Second, instead of mixing the posterior distribution with a single surrogate distribution, a multiple-component mixture formulation can be considered. Third, although the intuitive idea of first using some deterministic algorithm to find modes and then conducting MCMC to do mode jumping has been around, an efficient way of achieving the intended goal has not been formulated precisely. Our proposed MTM-enhanced jumping strategy, together with the WL weight adjustment, can help achieve the goal. It is particularly useful  to identify some specific classes of models where this general methodology is straightforward and effective to apply. 

\section*{Acknowledgements}
We thank Pierre Jacob for helpful discussions and suggestions. Some of the numerical examples in the paper are implemented based on the R package debiasedhmc \citep{heng2019unbiased}.

\bibliographystyle{chicago}
\bibliography{MTM.bib}

\end{document}